\documentclass[a4paper,12pt]{article}
\pdfoutput = 1
\usepackage{graphicx}
\usepackage{float}
\usepackage{subfigure}
\usepackage{mathrsfs}
\usepackage{amsmath}
\usepackage{amssymb}
\usepackage{setspace}
\usepackage{appendix}
\usepackage{multirow}
\usepackage{ctable}
\usepackage{epstopdf}
\usepackage{cite}
\usepackage{footnote}
\usepackage{hyperref}
\usepackage{mcite}
\usepackage{acronym}
\usepackage{authblk}

\newcommand{\znbb}{0\nu\beta\beta}
\newcommand{\ul}{\underline}
\newcommand{\mee}{\langle m_{ee}\rangle}
\newcommand{\sumnu}{\sum m_i}

\newcommand\T{\rule{0pt}{2.8ex}} 
\newcommand\B{\rule[-1.2ex]{0pt}{0pt}} 
\makesavenoteenv{tabular} 
\newcommand{\csection}[1]{\section*{\centering #1}}

\acrodef{GUT}{Grand Unified Theory}
\acrodef{PMNS}{Pontecorvo-Maki-Nakagawa-Sakata}
\acrodef{SM}{standard model}
\acrodef{TBM}{tribimaximal mixing}
\acrodef{VEV}{vacuum expectation value}
\acrodef{A-F}{Altarelli-Feruglio}
\acrodef{M-R}{Ma-Rajasekaran}

\title{\bf Deviations from tribimaximal mixing due to the vacuum expectation value misalignment in $A_4$ models}
\author{James Barry\footnote{E-mail: {\tt james.barry@mpi-hd.mpg.de}} \ and Werner Rodejohann\footnote{E-mail: {\tt werner.rodejohann@mpi-hd.mpg.de}}}
\date{}

\affil{Max-Planck-Institut f\"{u}r Kernphysik \\ Postfach 103980, D-69029 Heidelberg, Germany}
\begin{document}

\maketitle
\begin{abstract}
The addition of an $A_4$ family symmetry and extended Higgs sector to
the standard model can generate the tribimaximal mixing pattern for leptons,
assuming the correct vacuum expectation value alignment of the Higgs
scalars. Deviating this alignment affects the predictions for the
neutrino oscillation and neutrino mass observables. An attempt is made
to classify the plethora of models in the literature, with respect to
the chosen $A_4$ particle assignments. Of these models, two
particularly popular examples have been analyzed for deviations from
tribimaximal mixing by perturbing the vacuum expectation value alignments. The effect of
perturbations on the mixing angle observables is studied. 
However, it is only investigation of the mass-related observables (the
effective mass for neutrinoless double beta decay and the sum of
masses from cosmology) that can lead to the exclusion of particular
models by constraints from future data, which 
indicates the importance of neutrino mass in disentangling models. 
 The models have also been tested for fine-tuning of the parameters. Furthermore, a well-known seesaw model is generalized to include additional scalars, which transform as representations of $A_4$ not included in the original model.
\end{abstract}
\thispagestyle{empty}

\newpage

\section{Introduction} \label{sect:intro}

The experimental evidence of neutrino oscillations implies massive
neutrinos, which contradicts the predictions of the \ac{SM}. 
There are currently many experiments focused on precise 
measurements of the neutrino mass and mixing parameters: 
neutrino physics can be said to have entered the ``precision era''. 

Global fits to the latest neutrino oscillation data
\cite{Fogli:2005cq,Fogli:2006yq,Fogli:2008ig,Schwetz:2008osc} show
that the leptonic mixing, or Pontecorvo-Maki-Nakagawa-Sakata, 
matrix $U_{\rm PMNS}$ is very close to the \ac{TBM} matrix
\begin{equation}
  U_{\rm TBM} \equiv \begin{pmatrix} \frac{2}{\sqrt{6}} & \frac{1}{\sqrt{3}} & 0 \\ -\frac{1}{\sqrt{6}} & \frac{1}{\sqrt{3}} & \frac{1}{\sqrt{2}} \\ -\frac{1}{\sqrt{6}} & \frac{1}{\sqrt{3}} & -\frac{1}{\sqrt{2}} \end{pmatrix},
\label{eq:U_TBM}
\end{equation}
first proposed in Ref.~\cite{Harrison:2002tbm}. Since the allowed deviations from \ac{TBM} can only be small (not more than 10-15\%), this mixing pattern represents at least a zeroth order approximation to lepton mixing \cite{Pakvasa:2007zj}. It is completely different from the mixing in the quark sector, and has motivated extensive research into models of family symmetries \cite{Altarelli:2010gt}. 

Some of the discrete family symmetries used in the literature
are\footnote{See the review in Ref.~\cite{Altarelli:2010gt} for a
list of references.}: $A_4$, $S_3$, $S_4$, $T'$, $\Delta(27)$ and
$\Sigma(81)$; there are also models that employ continuous symmetries
such as $SU(3)$ or $SO(3)$. The $A_4$ models have a very economical
structure in terms of group representations and field content. 
The most general mass matrix leading to TBM can be shown to 
be invariant under one of the group generators
\cite{Altarelli:2010gt} (see the Appendix). 
Furthermore, the use of $A_4$ can be geometrically motivated: it is
the symmetry group of the regular tetrahedron, and the angle between
two faces is $2 \theta_{\rm TBM}$, where $\sin^2 \theta_{\rm TBM} =
\frac 13$. These characteristics have led many authors to construct
and/or study models based on $A_4$. Some models generate neutrino
masses via effective dimension-5 operators, some apply the type I
seesaw mechanism, whereas others use the type II, or the type I + II
seesaw mechanisms. Table \ref{table:a4_models_assign}\begin{savenotes}
\begin{table}[ht]
\centering
\caption[Particle assignments of $A_4$ models in the
literature]{Particle assignments of $A_4$ models in the
literature. Lepton doublets, charged lepton singlets and right-handed
neutrinos are denoted by $L_i$, $\ell^c_i$ and $\nu^c_i$,
respectively. $\Delta$ denotes the Higgs triplets in the type II
seesaw mechanism. Models that also study the quark sector have the superscript $^{\#}$, those that embed $A_4$ into a grand unified theory group have the superscript $^*$.}
\label{table:a4_models_assign}
\vspace{8pt}
\begin{tabular}{cccccc}
  \hline \hline Type  \T \B & $L_i$ & $\ell^c_i$ & $\nu^c_i$ & $\Delta$ & References \\
  \hline A1 \T & \multirow{2}{*}{$\ul{3}$} & \multirow{2}{*}{$\ul{1}$, $\ul{1}'$, $\ul{1}''$} & \multirow{2}{*}{$\cdots$} & $\cdots$ & \cite{Altarelli:2005a4,Zee:2005a4,Adhikary:2006wi,Altarelli:2005yx,Honda:2008rs,Brahmachari:2008a4,Feruglio:2008ht,Morisi:2009qa,Morisi:2009sz,Altarelli:2006kg} \cite{Bazzocchi:2007na}$^{\#}$ \\[0.8mm]
   A2 \T & & & & $\ul{1}$, $\ul{1}'$, $\ul{1}''$, $\ul{3}$ & \cite{Ma:2004zv,Ma:2005a42} \\[0.8mm]
   \hline B1 \T & \multirow{2}{*}{$\ul{3}$} & \multirow{2}{*}{$\ul{1}$, $\ul{1}'$, $\ul{1}''$} & \multirow{2}{*}{$\ul{3}$} & $\cdots$ & \cite{Altarelli:2005yx,Ma:2002yp,Babu:2002dz,Hirsch:2003dr,He:2006dk}$^{\#}$ \cite{Altarelli:2008bg,Burrows:2009pi}$^{*}$ \cite{Ma:2001a4,Babu:2005se,Ma:2005qf,Yin:2007rv,Adhikary:2008au,Csaki:2008qq,Chen:2009um,Hayakawa:2009va,Berger:2009tt,Ding:2009gh,Mitra:2009jj,delAguila:2010vg}  \\[0.8mm]  B2 \T & & & & $\ul{1}$, $\ul{3}$ & \cite{Dong:2010gk}$^{\#}$ \\[0.8mm]
   \hline C1 \T & \multirow{5}{*}{$\ul{3}$} & \multirow{5}{*}{$\ul{3}$} & \multirow{5}{*}{$\cdots$} & $\cdots$ & \cite{Zee:2005a4}\\[0.8mm]
   C2 \T & & & & $\ul{1}$ & \cite{Ma:2006vq,Ma:2006wm} \cite{Bazzocchi:2007au}$^{\#}$ \\[0.8mm]
   C3 \T & & & & $\ul{1}$, $\ul{3}$ & \cite{Ma:2009wi} \\[0.8mm]
   C4 \T & & & & $\ul{1}$, $\ul{1}'$, $\ul{1}''$, $\ul{3}$ & \cite{Hirsch:2005mc} \\[0.8mm]
   \hline D1 \T & \multirow{5}{*}{$\ul{3}$} & \multirow{5}{*}{$\ul{3}$} & \multirow{5}{*}{$\ul{3}$} & $\cdots$ & \cite{Morisi:2007ft,Ciafaloni:2009ub}$^*$ \cite{Hirsch:2008rp,Morisi:2009sc} \\[0.8mm]
   D2 \T & & & & $\ul{1}$ & \cite{Chen:2005jm} \cite{Bazzocchi:2008sp}$^{*}$ \\[0.8mm]
   D3 \T & & & & $\ul{1}'$ & \cite{Bazzocchi:2008rz}$^{*}$ \\[0.8mm]
   D4 \T & & & & $\ul{1}'$, $\ul{3}$ & \cite{Ciafaloni:2009qs}$^{*}$ \\[0.8mm]
   \hline E \T & $\ul{3}$ & $\ul{3}$ & $\ul{1}$, $\ul{1}'$, $\ul{1}''$ & $\cdots$ & \cite{Ma:2005a4,Lavoura:2006hb}\\[0.8mm]
   \hline F \T & $\ul{1}$, $\ul{1}'$, $\ul{1}''$ & $\ul{3}$ & $\ul{3}$ & $\ul{1}$ or $\ul{1}'$ & \cite{Hirsch:2007kh}  \\[0.8mm]
   \hline G \T & $\ul{3}$ & $\ul{1}$, $\ul{1}'$, $\ul{1}''$ & $\ul{1}$, $\ul{1}'$, $\ul{1}''$ & $\cdots$ & \cite{Frampton:2008ci} \\[0.8mm]
   \hline H \T & $\ul{3}$ & $\ul{1}$, $\ul{1}$, $\ul{1}$ & $\cdots$ & $\cdots$ & \cite{Lin:2008aj} \\[0.8mm]
   \hline I \T & $\ul{3}$ & $\ul{1}$, $\ul{1}$, $\ul{1}$ & $\ul{1}$, $\ul{1}$, $\ul{1}$ & $\cdots$ & \cite{King:2006np}$^{*}$ \\[0.8mm]
   \hline J \T & $\ul{3}$ & $\ul{1}$, $\ul{1}$, $\ul{1}$ & $\ul{3}$ & $\cdots$ & \cite{Altarelli:2009kr,Lin:2009bw} \\[0.8mm]
\hline \hline 
\end{tabular}
\end{table}
\end{savenotes} is an attempt to
classify the vast number of models,\footnote{An earlier, much less
complete classification can be found in \cite{Morisi:2008nk}.}
according to the chosen $A_4$ assignment of the lepton doublets,
lepton singlets and, if appropriate, the seesaw
particles.\footnote{There are also models that use the inverse and
linear seesaw mechanisms \cite{Hirsch:2009mx}, as well as the inverse
type III seesaw mechanism \cite{Ibanez:2009du}, with the same particle
assignments as type D models.} The majority fall into the first four categories.

Very often the \ac{TBM} scheme is obtained only approximately, or with the cost of fine-tuning and/or various assumptions, such as \acf{VEV} alignment. These alignments are chosen, or the models are explicitly constructed, in order to reach alignment, resulting in a certain mixing pattern (in this case \ac{TBM}). However, corrections to the \ac{VEV} alignment are expected, be it from renormalization, higher order operators, or the tree-level exchange of heavy fermions, for example. The aim of this paper is to study the effects of
\ac{VEV}-misalignment on the neutrino mass and lepton mixing observables. There already exist some numerical analyses \cite{Honda:2008rs,Brahmachari:2008a4,Adhikary:2008au,Lavoura:2006hb,Altarelli:2009kr} 
focused on specific $A_4$ models. In addition, the effects of higher order operators have been studied in $A_4$ \cite{Altarelli:2009kr} and $S_4$ \cite{Ding:2009iy,Bazzocchi:2009pv} models, where the unperturbed \ac{VEV} alignments predict exact \ac{TBM}.  This work emphasizes that observables related to neutrino mass (that is, the effective mass for neutrinoless double beta decay ($\znbb$) and the sum of neutrino masses for cosmology) provide the best possibility to disentangle the models. Furthermore, and in contrast to previous studies, a more general \ac{VEV}-misalignment is allowed for. The analysis in the present paper is focused on models of types A and B that predict \ac{TBM}, as well as generalizations of these models to include more Higgs singlets.

In this analysis,\footnote{Other approaches to deviations from TBM can be found in Refs.~\cite{Plentinger:2005kx,Dighe:2006sr,Hirsch:2006je,Hochmuth:2007wq,Albright:2008tbm,Goswami:2009yy,Li:2009kx,Ge:2010js}.} the chosen VEV alignment is modified by random complex deviations, perturbing the neutrino and charged lepton mass matrices from their original structure ($M_{\nu}$ and $M_{\ell}$) to the perturbed ones, $M'_{\nu}$ and $M'_{\ell}$. The resulting neutrino mixing angles and mass-squared differences can be compared with current data (Table~\ref{table:oscparameters}).
\begin{table}[ht]
  \centering
   \caption[Best-fit values and allowed $n\sigma$ ranges for the global three flavor neutrino oscillation parameters]{Best-fit values and allowed $n\sigma$ ranges for the global three flavor neutrino oscillation parameters, from Ref.~\cite{Fogli:2008ig}.}
\label{table:oscparameters}
\vspace{2mm}
  \begin{footnotesize}
  \begin{tabular}{cccccc}
    \hline \hline \T\B Parameter & $\Delta m_{21}^2 \ (10^{-5}\, {\rm eV}^2)$ & $\sin^2\theta_{12}$ & $\sin^2\theta_{13}$ & $\sin^2\theta_{23}$ & $|\Delta m_{31}^2| \ (10^{-3}\, {\rm eV}^2)$ \\
    \hline \T Best fit & $7.67$ & $0.312$ & $0.016$ & $0.466$ & $2.39$ \\
    $1\sigma$ range & $7.48\!-\!7.83$ & $0.294\!-\!0.331$ & $0.006\!-\!0.026$ & $0.408\!-\!0.539$ & $2.31\!-\!2.50$ \\
    $2\sigma$ range & $7.31\!-\!8.01$ & $0.278\!-\!0.352$ & $<\!0.036$ & $0.366\!-\!0.602$ & $2.19\!-\!2.66$ \\
    $3\sigma$ range & $7.14\!-\!8.19$ & $0.263\!-\!0.375$ & $<\!0.046$ & $0.331\!-\!0.644$ & $2.06\!-\!2.81$ \\[1.5mm]
    \hline \hline
   \end{tabular}
   \end{footnotesize}
\end{table} 
The well-known standard parameterization of the 
PMNS mixing matrix is 
\begin{equation}
  U_{\rm PMNS} = \begin{pmatrix}
    c_{12}c_{13} & s_{12}c_{13} & s_{13}e^{-i\delta} \\
    -s_{12}c_{23} - c_{12}s_{23}s_{13}e^{i\delta} & c_{12}c_{23} - s_{12}s_{23}s_{13}e^{i\delta} & s_{23}c_{13} \\
    s_{12}s_{23} - c_{12}c_{23}s_{13}e^{i\delta} & -c_{12}s_{23} - s_{12}c_{23}s_{13}e^{i\delta} & c_{23}c_{13}
    \end{pmatrix}
    \begin{pmatrix} 1 & 0 & 0 \\ 0 & e^{i\lambda_{2}} & 0 \\ 0 & 0 &
e^{i\lambda_{3}} \end{pmatrix} , 
\label{eq:PMNSmatrix}
\end{equation}
where $c_{ij} \equiv \cos\theta_{ij}$, $s_{ij} \equiv \sin\theta_{ij}$ and $\theta_{12}, \theta_{13}, \theta_{23}$ ($0 \leq \theta_{ij} \leq \pi/2$) are the three mixing angles. 
There are three phases in Eq.~\eqref{eq:PMNSmatrix}: $\delta$ is the {\em CP} violating Dirac phase, and $\lambda_{2}$ and $\lambda_{3}$ are Majorana phases, with $0 \leq \delta,\lambda_2,\lambda_3 \leq 2\pi$. The two Majorana phases, $\lambda_2$ and $\lambda_3$, do not affect the neutrino oscillation probability, but have an influence on the amplitude for $\znbb$.

One can also perform a ``fine-tuning test'' for each model, by examining the values that the mass matrix parameters must take in order to give the correct mass-squared differences, before perturbations are applied. Since these parameters generally originate from the product of some coupling constant with the \ac{VEV} of a Higgs scalar, any close relationship between the parameters is highly unlikely, and could be evidence of fine-tuning in a particular model \cite{Brahmachari:2008a4}. \\

The paper is built up as follows: in Section~\ref{sect:typeA} a type A model is introduced, it is examined for fine-tuning, the addition of Higgs singlets is discussed, and the model is analyzed for deviations from \ac{TBM}; in Section~\ref{sect:AFSSmodel} the same procedure is followed for a type B seesaw model. Section~\ref{sect:conclusion} presents the summary and conclusions, and for the sake of completeness there is a discussion of the $A_4$ group in the Appendix.

\section{The original Ma/Altarelli-Feruglio type A model} \label{sect:typeA}

In type A models, lepton doublets transform as $\ul{3}$, charged
lepton singlets as $\ul{1}, \ul{1}', \ul{1}''$, and right-handed
neutrinos are absent. In this case the neutrino mass usually comes
from dimension-5 operators. Although
Table~\ref{table:a4_models_assign} contains a long list of references
for type~A models, many of these works are phenomenological analyses
of the same few models. The original model by Ma \cite{Ma:2004zv} is
further developed in Ref.~\cite{Altarelli:2005a4}, where also an extra-dimensional solution to the vacuum alignment problem is provided.\footnote{Note that the model in Ref.~\cite{Ma:2004zv} contains 6 Higgs triplets, whereas the model in Ref.~\cite{Altarelli:2005a4} uses dimension-5 operators.}

The models in Refs.~\cite{Ma:2004zv}~and~\cite{Altarelli:2005a4} employ the so-called \ac{M-R} basis for $A_4$, in which neither $M_\nu$ nor $M_\ell$ is diagonal, but the product of the mixing matrices in each sector leads to \ac{TBM}. In order to connect $A_4$ models with the modular symmetry and thus the larger framework of string theory, the same model can be formulated \cite{Altarelli:2005yx} in a different basis for $A_4$ (the \ac{A-F} basis). In this basis the charged leptons immediately come out as diagonal, which means that the neutrino mass matrix is in the flavor basis, and is diagonalized by the \ac{TBM} matrix. The two bases are simply related by a unitary transformation, and the multiplication rules differ (see the Appendix for details).

\subsection{The original model in the \ac{A-F} basis} \label{sect:AFmodel}

Along with the usual type A particle assignments for leptons (Table~\ref{table:a4_models_assign}), this model has two SM Higgs doublets, which are invariant under $A_4$, as well as two $A_4$ triplets $\varphi$ and $\varphi'$, and an $A_4$ singlet $\xi$, all three of which are gauge singlets (Table~\ref{table:AFmodel_assign}).
\begin{table}[t]
\centering
\caption{Particle assignments of the A-F $A_4$ model. There is also an
additional $Z_3$ symmetry, which decouples the charged lepton and
neutrino sectors, and a $U(1)$ symmetry to generate the hierarchy of charged
lepton masses.}
\label{table:AFmodel_assign}
\vspace{8pt}
\begin{tabular}{ccc}
  \hline \hline \T \B Lepton & $SU(2)_L$ & $A_4$ \\
  \hline \T $L$ & $2$ & $\ul{3}$ \\
     $e^c$ & $1$ & $\ul{1}$ \\
     $\mu^c$ & $1$ & $\ul{1}''$ \\
   $\tau^c$ & $1$ & $\ul{1}'$ \\
   \hline \T\B Scalar & & \\
   \hline \T $h_u$ & $2$ & $\ul{1}$ \\
   $h_d$ & $2$ & $\ul{1}$ \\
   $\varphi$ & $1$ & $\ul{3}$ \\
   $\varphi'$ & $1$ & $\ul{3}$ \\
   $\xi$ & $1$ & $\ul{1}$ \\[1mm] \hline \hline
\end{tabular}
\end{table}
These particle assignments, along with the $A_4$ multiplication rules, lead to the Lagrangian
\begin{align}
  \mathscr{L}_{\rm Y} &= \ y_ee^c(\varphi L) + y_{\mu}\mu^c(\varphi L)' + y_{\tau}\tau^c(\varphi L)'' +  x_a\xi(LL) + x_d(\varphi'LL)  \notag \\[2mm]
  &[+\ x_c\xi'(LL)'' + x_b\xi''(LL)'] + {\rm H.c.} + \dots \ , 
\label{eq:lag_AFmodelnew}
\end{align}
where $(\ul{3}\ul{3})$ transforms as $\ul{1}$, $(\ul{3}\ul{3})'$ transforms as $\ul{1}'$, and $(\ul{3}\ul{3})''$ transforms as $\ul{1}''$, and $y_\alpha$, $x_a$ and $x_d$ are dimensionless coupling constants. The notation in Eq.~\eqref{eq:lag_AFmodelnew} follows the simplified description from Ref.~\cite{Altarelli:2005yx}, where the Higgs doublet fields $h_u$ and $h_d$, and the cut-off scale $\Lambda$ are set to 1. Thus the term $y_ee^c(\varphi L)$ is in fact $y_ee^c(\varphi L)h_d/\Lambda$, $x_a\xi(LL)$ is short for $x_a\xi(L h_uL h_u)/\Lambda^2$ and so on. The dots stand for higher dimensional operators -- in this model these are suppressed by additional powers of the cut-off $\Lambda$, as long as the VEVs are sufficiently smaller than $\Lambda$. The two terms in parenthesis on the second line of Eq.~\eqref{eq:lag_AFmodelnew} come from additional Higgs singlets; these were not part of the original model, but one can show \cite{Brahmachari:2008a4} that \ac{TBM} can still be achieved with either two or three Higgs singlets in this model. This will be discussed in Sections~\ref{sect:af2higgs}~and~\ref{sect:af3higgs}.

Upon symmetry breaking, the VEVs of the Higgs singlet and triplets take the alignments
\begin{equation}
  \langle\xi\rangle = u_a\ , \quad \langle\varphi\rangle = (v,0,0) \quad {\rm and} \quad \langle\varphi'\rangle = (v',v',v')\ ,
\label{eq:AFmodel_Higgsalign}
\end{equation}
which lead to the charged lepton mass matrix
\begin{equation}
  M_\ell = v_d\frac{v}{\Lambda} \begin{pmatrix} y_e & 0 & 0 \\ 0 & y_\mu & 0 \\ 0 & 0 & y_\tau \end{pmatrix}\ ,
\label{eq:mlep_AF}
\end{equation}
where $v_d$ is the \ac{VEV} of the Higgs doublet $h_d$. Thus the charged fermion masses are
\begin{equation}
  m_e = y_ev_d\frac{v}{\Lambda} \ , \quad m_{\mu} = y_{\mu}v_d\frac{v}{\Lambda} \ , \quad m_{\tau} = y_{\tau}v_d\frac{v}{\Lambda}\ .
\label{eq:lepmassesaf}
\end{equation}
\begin{figure}[ht]
 \centering 
\includegraphics[width=.6\textwidth]{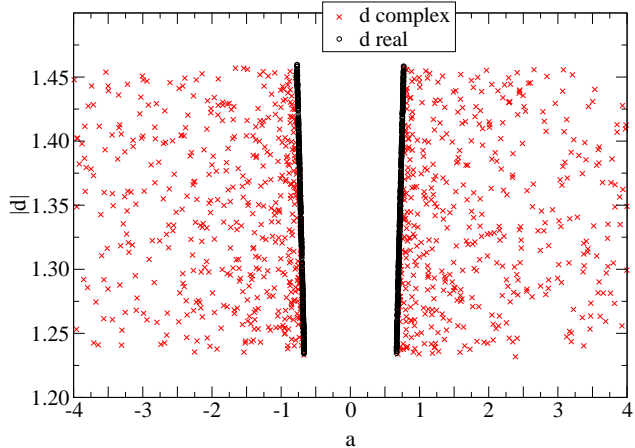}
\caption[Scatter plot of the $a - d$ parameter space in the A-F model, with one Higgs singlet, normal hierarchy]{Scatter plot showing allowed regions (for the $3\sigma$ ranges of the oscillation parameters) in the $a - d$ parameter space for the original A-F model, with one Higgs singlet, normal hierarchy and \ac{TBM}.}
\label{fig:avb_AF1Higgs_plot}
\end{figure}

When only one Higgs singlet ($\xi \sim \ul{1}$) is present, the neutrino mass matrix is
\begin{equation}
  M^{(1)}_\nu = m_0 \begin{pmatrix} a + \frac{2d}{3} & -\frac{d}{3} & -\frac{d}{3} \\[1mm] \cdot & \frac{2d}{3} & a - \frac{d}{3} \\[1mm] \cdot & \cdot & \frac{2d}{3} \end{pmatrix}\ ,
\label{eq:mnu_AF1H}
\end{equation}
with $m_0 = \frac{v_u^2}{\Lambda}$, $a = 2x_a\frac{u_a}{\Lambda}$ and $d = 2x_d\frac{v'}{\Lambda}$, where $v_u$ is the \ac{VEV} of $h_u$. The neutrino mass matrix is diagonalized by the transformation
\begin{equation}
  U^TM_{\nu}U = \frac{v_{u}^2}{\Lambda}\,{\rm diag}(a+d,a,-a+d)\ ,
\label{eq:afdiag}
\end{equation}
with $U = U_{\rm TBM}$, as in Eq.~\eqref{eq:U_TBM}. Thus \ac{TBM} is achieved, and the neutrino masses are $m_1 = m_0(a+d)$, $m_2 = m_0a$ and $m_3 = m_0(-a+d)$, which results in the sum-rule $2m_2 + m_3 = m_1$. Here the masses are understood to be complex, with the Majorana phases still attached. Note that with only one Higgs singlet it is impossible to get the inverted mass hierarchy in this model, as shown in Ref.~\cite{Brahmachari:2008a4}. 

It is interesting to note that in the case of one Higgs singlet, with
the mass matrix in Eq.~\eqref{eq:mnu_AF1H}, some fine-tuning is
required between the parameters $a$ and $d$ for the model to give the
correct neutrino mass-squared differences
\cite{Brahmachari:2008a4}. This seems rather contrived, since $a$ and
$d$ come from the products of different Yukawa couplings with the VEVs
of the Higgs singlet $\xi$ and triplet $\varphi'$, respectively. As
can be seen in Fig.~\ref{fig:avb_AF1Higgs_plot}, if both $a$ and $d$
are real (as in Ref.~\cite{Brahmachari:2008a4}), there is a linear
relationship between the two parameters. If $d$ is complex (as in this
analysis), there is only a slightly greater allowed region in the $a -
d$ parameter space. Note that w.l.o.g., $a$ can be chosen to be
real. There are no perturbations applied in this case, and the
parameter $m_0$ is set to $0.025$ eV, the typical scale for the
mass matrix of normally ordered neutrinos. In later cases, where the inverted mass ordering is studied, $m_0$ is fixed to 0.05 eV. 
The magnitudes of the parameters $a$ and $d$ (and later also $c$)  
are randomly varied in the range $|a,c,d| \leq 4$, with their complex 
phases varying from zero to $2\pi$. 

\subsection{Two Higgs singlets} \label{sect:af2higgs}

\begin{figure}[ht]
  \centering
  \subfigure[Normal hierarchy]{\label{fig:cd_nh_AF2H}
  \includegraphics[width=.48\textwidth]{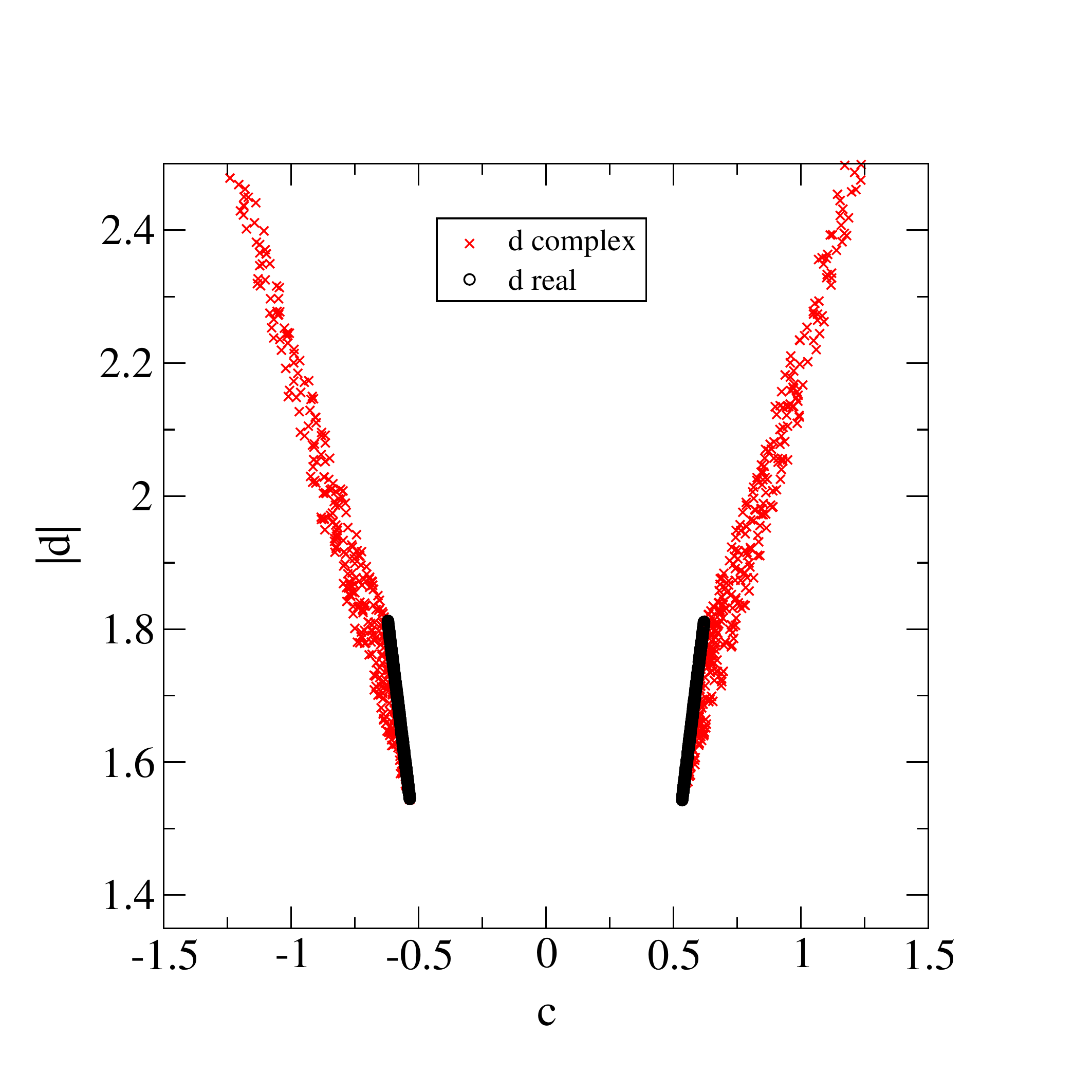}}
  \subfigure[Inverted hierarchy]{\label{fig:cd_ih_AF2H}
  \includegraphics[width=.48\textwidth]{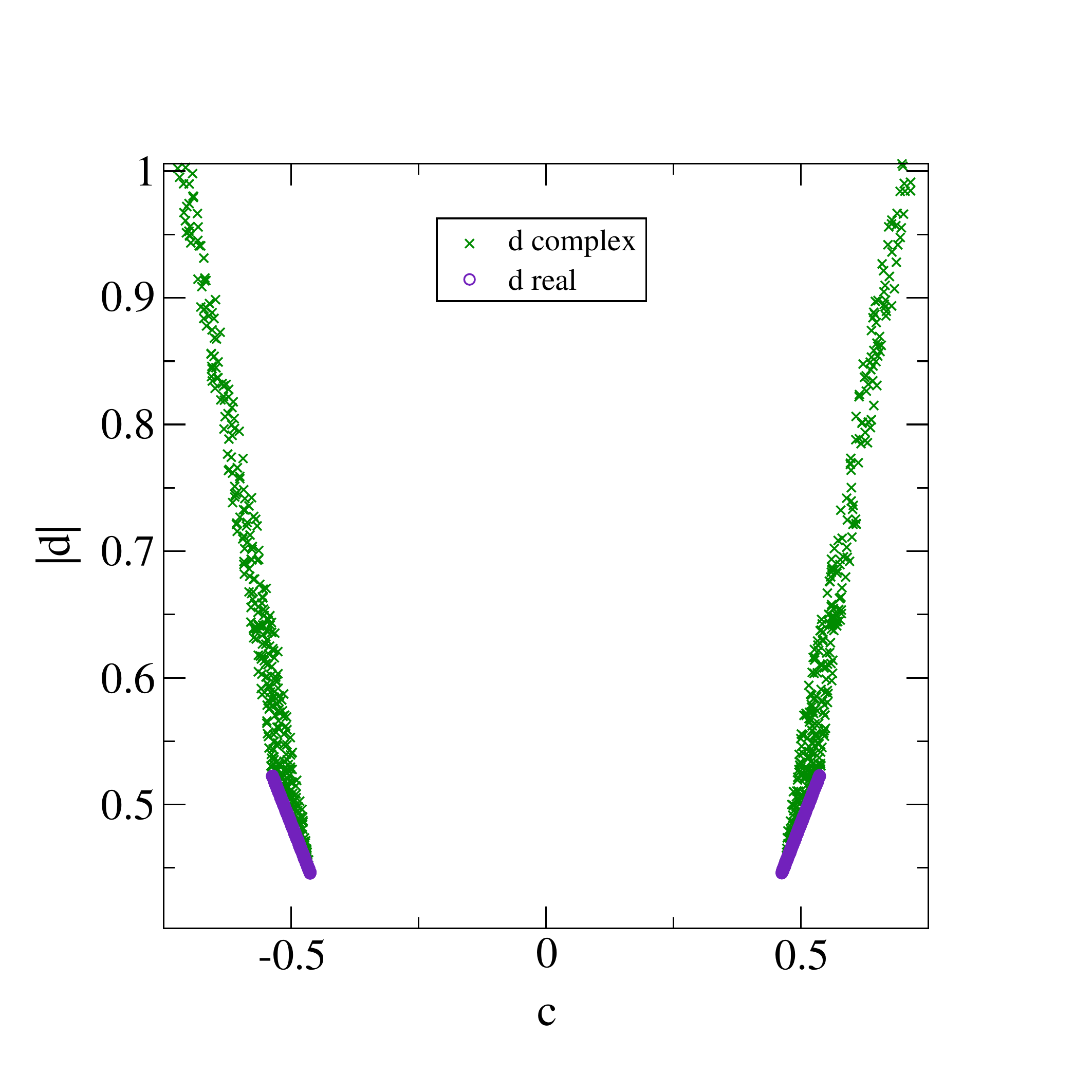}}
  \caption[Scatter plots of the $c-d$ parameter space for the A-F model with two Higgs singlets]{Scatter plots of the $c-d$ parameter space for the A-F model with two Higgs singlets [Eq.~\eqref{eq:mnu_AF2H}], for normal and inverted hierarchy, with the condition $b=c$. In order to emphasize the difference between the complex and real case, the entire parameter space is not shown: in the complex case $|d|$ ranges up to $4$ for the normal hierarchy and $2.5$ for the inverted hierarchy.}
  \label{fig:massparam_AF2H}
\end{figure}

Recall that only one Higgs singlet is introduced in the original model (Table~\ref{table:AFmodel_assign}). However, in the framework of $A_4$ symmetry it is natural to take advantage of all representations of the group, and in this model it is also possible to achieve \ac{TBM} with both two and/or three Higgs singlets \cite{Brahmachari:2008a4}. In addition to the Higgs singlet $\xi$, the singlets $\xi'$ and $\xi''$ can be introduced [Eq.~\eqref{eq:lag_AFmodelnew}], transforming as $\ul{1}'$ and $\ul{1}''$ under $A_4$, respectively. The new singlets have the \acp{VEV} 
\begin{equation}
  \langle \xi' \rangle = u_c \quad {\rm and} \quad \langle \xi'' \rangle = u_b\ .
\end{equation}
With only two Higgs singlets, there are three possible combinations ($\xi$,~$\xi'$; $\xi$,~$\xi''$ and $\xi'$,~$\xi''$), but one can show \cite{Brahmachari:2008a4} that only the singlets $\xi'$ and $\xi''$ can give rise to \ac{TBM}. In this case, the resulting mass matrix is 
\begin{equation}
  M^{(2)}_\nu = m_0 \begin{pmatrix} \frac{2d}{3} & b-\frac{d}{3} & c-\frac{d}{3} \\[1mm] \cdot & c+\frac{2d}{3} & -\frac{d}{3} \\[1mm] \cdot & \cdot & b+\frac{2d}{3} \end{pmatrix}\ ,
\label{eq:mnu_AF2H}
\end{equation}
where $b = 2x_b\frac{u_b}{\Lambda}$ and $c =
2x_c\frac{u_c}{\Lambda}$. An additional condition for \ac{TBM} is that
$b = c$, which is a consequence of the necessary $\mu-\tau$ symmetry,\footnote{Although $\mu-\tau$ symmetry is required to get \ac{TBM}, this forces one to impose the {\em ad hoc} relation $b = c$.}
and with this constraint the eigenvalues turn out to be $m_1 =
m_0(-c+d)$, $m_2 = 2m_0c$ and $m_3 = m_0(c+d)$, with the new sum-rule
$m_3-m_1=m_2$. In this case, w.l.o.g., $c$ can be chosen to be
real. The scatter plots in Fig.~\ref{fig:massparam_AF2H} show that the
$c-d$ parameter space is quite tightly constrained (note that with
additional Higgs singlets, the inverted mass hierarchy is now
possible). 

\subsection{Three Higgs singlets} \label{sect:af3higgs}

\begin{figure}[ht]
  \centering
  \subfigure[Normal hierarchy]{\label{fig:acd_nh_AF3H}
  \includegraphics[width=.75\textwidth]{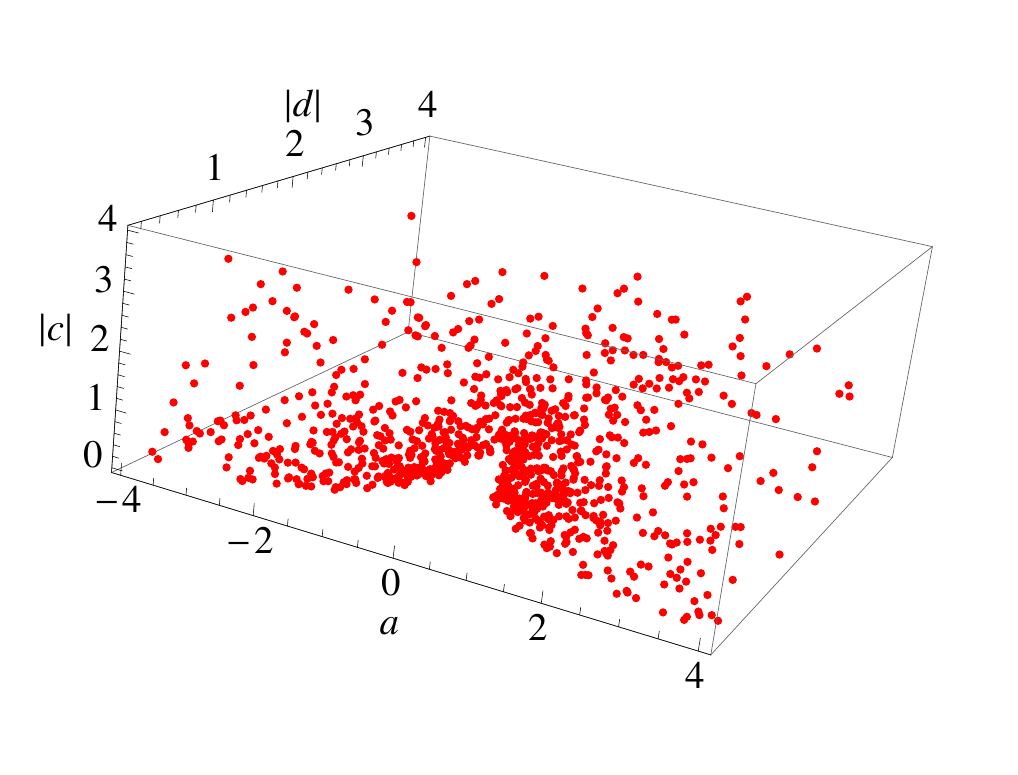}} 
  \subfigure[Inverted hierarchy]{\label{fig:acd_ih_AF3H}
  \includegraphics[width=.75\textwidth]{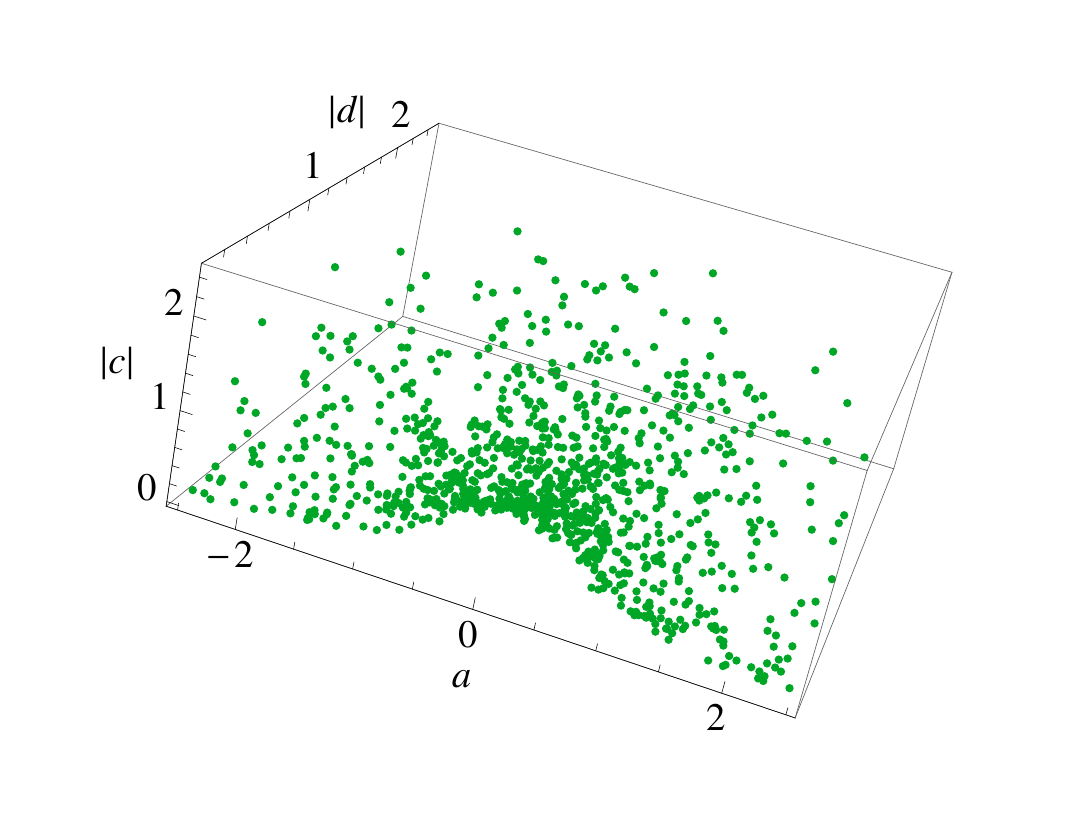}}
  \caption[Scatter plots of the $a-c-d$ parameter spaces for the A-F model with three Higgs singlets]{Scatter plots of the $a-c-d$ parameter spaces for the A-F model with three Higgs singlets [Eq.~\eqref{eq:mnu_AF3H}], for normal and inverted hierarchy, with the condition $b=c$.}
  \label{fig:massparam_AF3H}
\end{figure}

If all three singlets ($\xi$, $\xi'$ and $\xi''$) are present, the resulting mass matrix is
\begin{equation}
  M^{(3)}_\nu = m_0 \begin{pmatrix} a + \frac{2d}{3} & b-\frac{d}{3} & c-\frac{d}{3} \\[1mm] \cdot & c+\frac{2d}{3} & a - \frac{d}{3} \\[1mm] \cdot & \cdot & b+\frac{2d}{3} \end{pmatrix}\ ,
\label{eq:mnu_AF3H}
\end{equation}
and the requirement for exact \ac{TBM} is that $a \neq b = c$, which
again reflects the necessary $\mu-\tau$ symmetry.\footnote{It can be
shown \cite{Brahmachari:2008a4} that the conditions $a = b = c$, $a =
b \neq c$ and $a = c \neq b$ do not simultaneously give \ac{TBM} and
the correct mass spectrum.} Here one can choose real $a$ and complex
$c$ and $d$, w.l.o.g. This case is equivalent to the original Ma model
in Ref.~\cite{Ma:2004zv}, and here there is more freedom in choosing
parameters, as can be seen from the scatter plots of $a-c-d$ parameter
space in Fig.~\ref{fig:massparam_AF3H}. There is basically no more
tuning necessary in order to generate the correct mass-squared differences. 
The eigenvalues of the mass matrix in Eq.~\eqref{eq:mnu_AF3H}, with $a \neq b = c$, are $m_1 = m_0(a-c+d)$, $m_2 = m_0(a+2c)$ and $m_3 = m_0(-a+c+d)$.

\subsection{Deviations from \ac{TBM} in the A-F model} \label{sect:devaf}

The three mass matrices in Eqs.~\eqref{eq:mnu_AF1H}, \eqref{eq:mnu_AF2H} and \eqref{eq:mnu_AF3H} are phenomenologically interesting, and will be numerically analyzed below. 
In order to study deviations from \ac{TBM}, the VEV alignments of the Higgs triplets are perturbed, so that
\begin{equation}
  \langle\varphi\rangle = (v,v\,\epsilon^{\rm ch}_1,v\,\epsilon^{\rm ch}_2) \quad {\rm and} \quad \langle\varphi'\rangle = (v',v'(1+\epsilon_1),v'(1+\epsilon_2))\ .
\label{eq:AF_VEVdev}
\end{equation}
Furthermore, in the cases of two and three Higgs singlets, 
\begin{equation} \label{eq:bc}
b = c\,(1 + \epsilon_3)
\end{equation}
is defined in order to study the effect of changing the relative alignment of the Higgs singlets. Recall that the condition $b=c$ is necessary for \ac{TBM} in both the two and three singlet cases. 

With the above \ac{VEV}-misalignment, the charged lepton mass matrix becomes
\begin{equation}
  M'_\ell = v_d\frac{v}{\Lambda} \begin{pmatrix} y_e & y_e\,\epsilon^{\rm ch}_2 & y_e\,\epsilon^{\rm ch}_1 \\ y_\mu\,\epsilon^{\rm ch}_1 & y_\mu & y_\mu\,\epsilon^{\rm ch}_2 \\ y_\tau\,\epsilon^{\rm ch}_2 & y_\tau\,\epsilon^{\rm ch}_1 & y_\tau \end{pmatrix} .
\label{eq:mlepdev_AF}
\end{equation}
In the unperturbed case, the mass of each charged lepton $l_\alpha$ is
$m_\alpha = y_\alpha v_d\frac{v}{\Lambda}$
[Eq.~\eqref{eq:lepmassesaf}]. In this analysis, the mass scale
$v_d\frac{v}{\Lambda}$ is fixed to the tau mass, and each of the
coefficients $y_e$, $y_\mu$ and $y_\tau$ are varied randomly by 10\%
around their unperturbed values. Note that the charged lepton sector is unaffected by additional Higgs singlets, due to the presence of a $Z_3$ symmetry (Table~\ref{table:AFmodel_assign}).
\begin{figure}[ht]
  \centering
  \subfigure[]{\label{fig:sinsq1323_AF_n}
  \includegraphics[width=.5\textwidth]{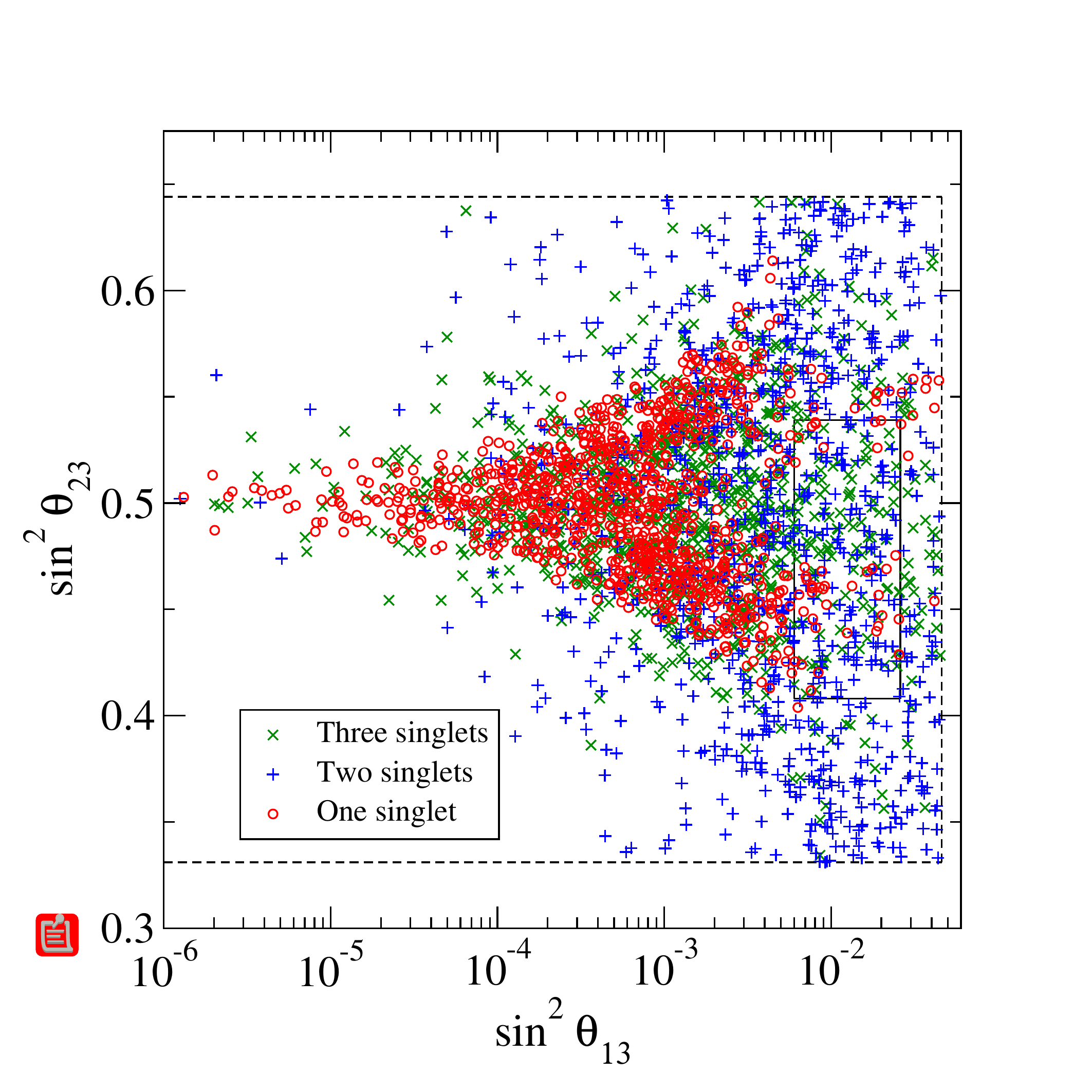}} \hspace{-1cm}
  \subfigure[]{\label{fig:JCP_AF_n}
  \includegraphics[width=.52\textwidth]{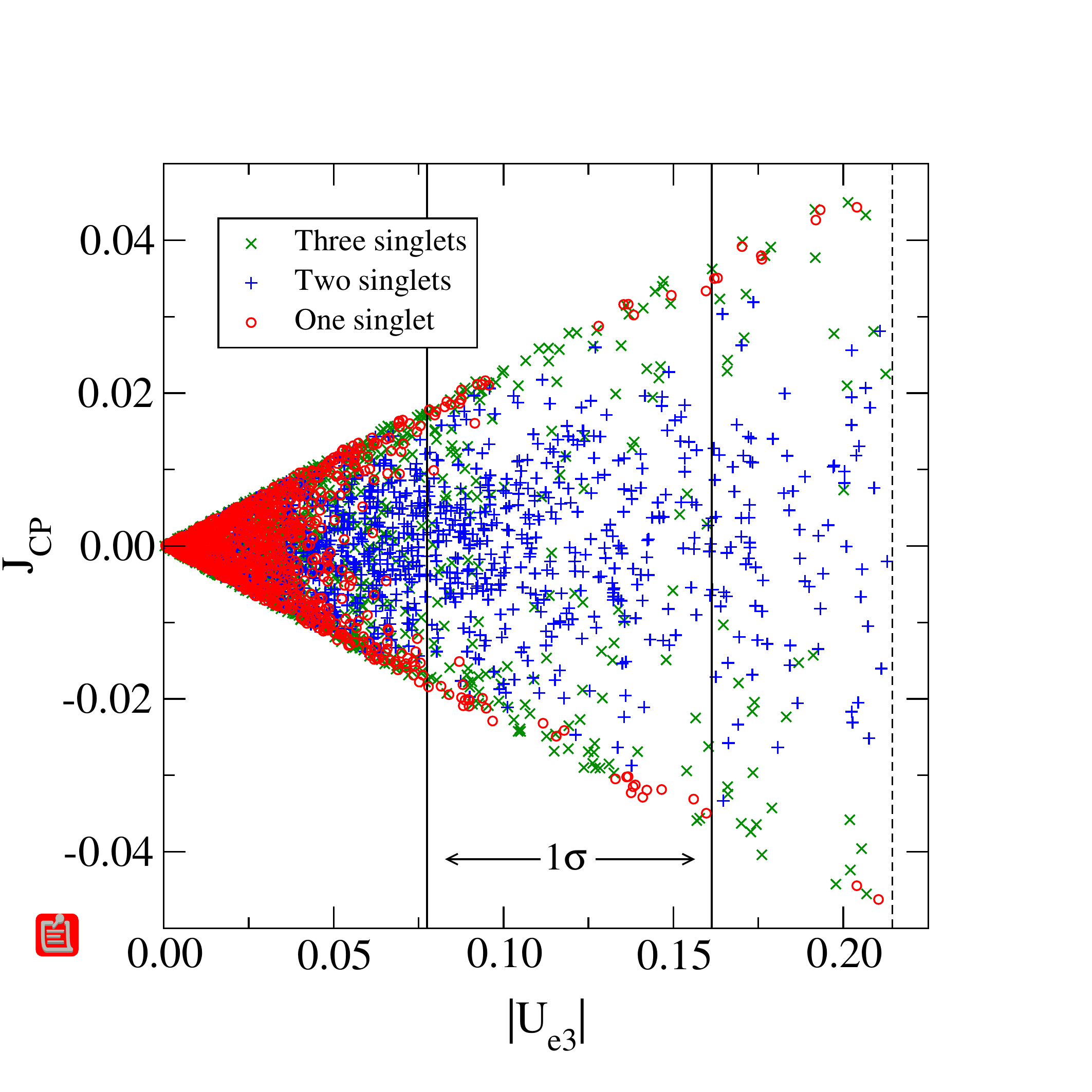}}
  \caption[Scatter plots of mixing angles and $J_{\rm CP}$ for the A-F model, normal hierarchy, with different numbers of Higgs singlets]{Scatter plots of (a) $\sin^2\theta_{23}$ against $\sin^2\theta_{13}$ and (b) $J_{\rm CP}$ against $|U_{e3}|$ for the A-F model, normal hierarchy, with one (red circles), two (blue plus signs) and three (green crosses) Higgs singlets. The solid and dashed lines denote the $1\sigma$ and $3\sigma$ allowed regions, respectively.}
  \label{fig:sinsq1323_JCP_AF_n}
\end{figure}

The deviated neutrino mass matrix with one Higgs singlet is 
\begin{equation}
  {M^{(1)}_\nu}' = m_0\begin{pmatrix} a + \frac{2d}{3} & - \frac{d}{3}(1 + \epsilon_2) & - \frac{d}{3}(1 + \epsilon_1) \\[1mm] \cdot & \frac{2d}{3}(1 + \epsilon_1) & a - \frac{d}{3} \\[1mm] \cdot & \cdot & \frac{2d}{3}(1 + \epsilon_2)\end{pmatrix}\ ,
\label{eq:mnudev_AF1H}
\end{equation}
and with two Higgs singlets ($\xi'$ and $\xi''$) is
\begin{equation}
  {M^{(2)}_\nu}' = m_0\begin{pmatrix} \frac{2d}{3} & c\,(1 + \epsilon_3) - \frac{d}{3}(1 + \epsilon_2) & c - \frac{d}{3}(1 + \epsilon_1) \\[1mm] \cdot & c+\frac{2d}{3}(1 + \epsilon_1) & - \frac{d}{3} \\[1mm] \cdot & \cdot & c\,(1 + \epsilon_3)+\frac{2d}{3}(1 + \epsilon_2)\end{pmatrix}\ .
\label{eq:mnudev_AF2H}
\end{equation}
The most general case (three Higgs singlets) is
\begin{equation}
  {M^{(3)}_\nu}' = m_0\begin{pmatrix} a + \frac{2d}{3} & c\,(1 + \epsilon_3) - \frac{d}{3}(1 + \epsilon_2) & c - \frac{d}{3}(1 + \epsilon_1) \\[1mm] \cdot & c+\frac{2d}{3}(1 + \epsilon_1) & a - \frac{d}{3} \\[1mm] \cdot & \cdot & c\,(1 + \epsilon_3)+\frac{2d}{3}(1 + \epsilon_2)\end{pmatrix} \ ,
\label{eq:mnudev_AF3H}
\end{equation}
where the condition $a \neq c$ still holds. One proceeds by
diagonalizing the matrices in Eqs.~\eqref{eq:mnudev_AF1H},
\eqref{eq:mnudev_AF2H} and \eqref{eq:mnudev_AF3H}. The perturbation
parameters are in general complex, and the range $|\epsilon_i^{\rm (ch)}|
\leq 0.3$ is used throughout this work, with the phases varied freely.
W.l.o.g., one can choose the parameters 
$\epsilon_1$ and $\epsilon^{\rm ch}_1$ to be real. 
The other parameters $a$, $b$ and $d$ are varied as before, and $m_0$ is
also fixed as described above. 

It is interesting to compare the deviations from \ac{TBM} for
different numbers of Higgs singlets, with the same perturbations
applied to $M_\ell$ in each case
[Eq.~\eqref{eq:mlepdev_AF}]. Fig.~\ref{fig:sinsq1323_JCP_AF_n} shows
the results for the normal mass hierarchy (it is impossible to get the
inverted hierarchy with one Higgs singlet). 
There are small differences, and in general one can conclude that with more singlets, greater deviation from \ac{TBM} is possible. However, it is evident that if \ac{VEV} alignment deviations are applied, the $A_4$ models deviate from \ac{TBM} in a rather random fashion, and it is difficult to draw any firm conclusions from the plots of mixing angle observables.
\begin{figure}[ht]
  \centering 
  \includegraphics[width=0.8\textwidth]{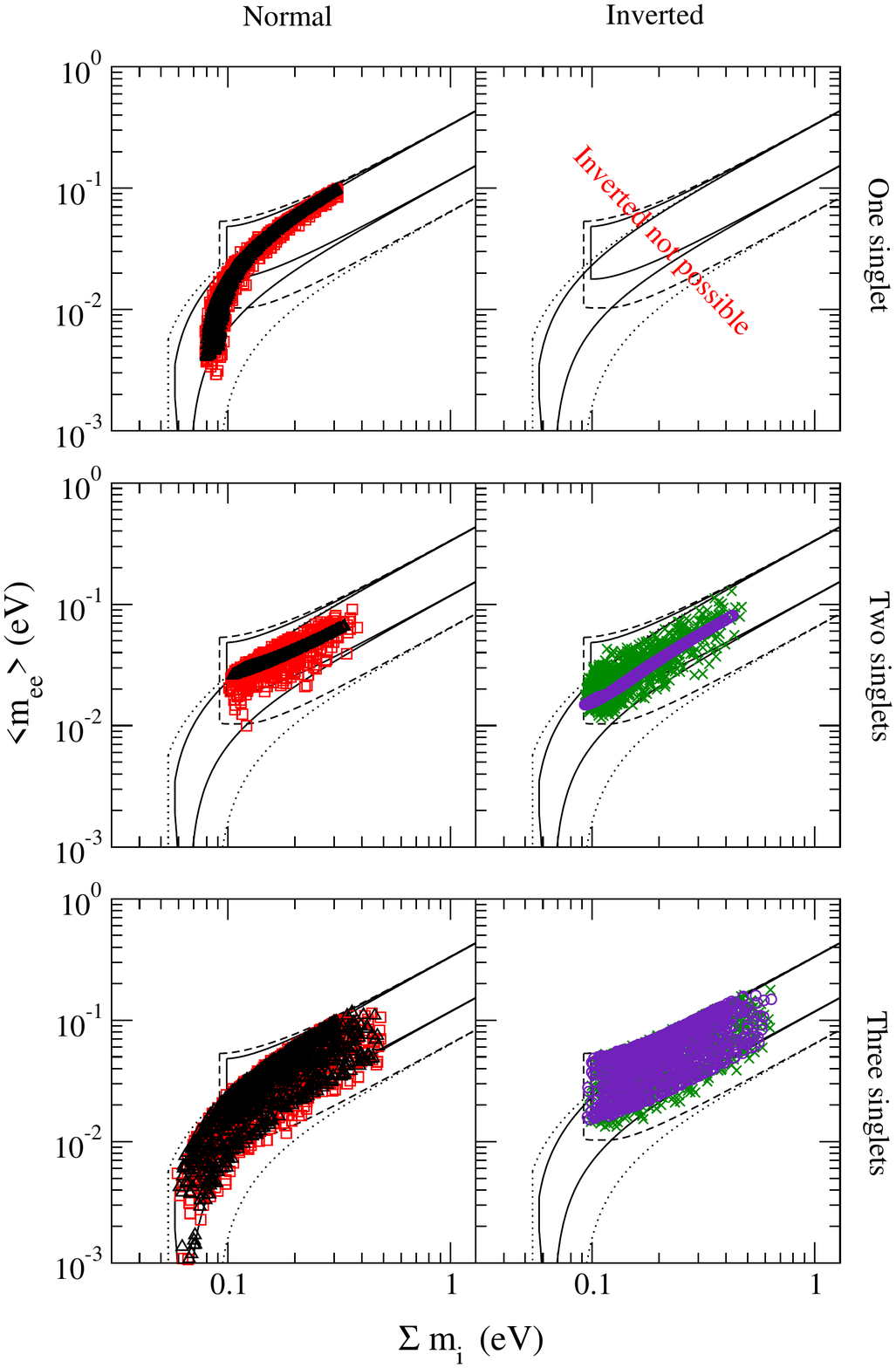}
		\caption[Scatter plots of mass observables for the A-F model, with one, two and three Higgs singlets]{Scatter plots of $\mee$ against $\sumnu$ for the original A-F model, with one, two and three Higgs singlets. Black triangles (red squares) denote the normal hierarchy unperturbed (perturbed) case; indigo circles (green crosses) denote the inverted hierarchy unperturbed (perturbed) case.}
 \label{fig:mAFSand_plot}
\end{figure}

In contrast, the mass-dependent observables $\sumnu$ (the sum of absolute neutrino masses) and $\mee$ (the effective mass for $\znbb$) allow for comparison between the three cases presented above (Fig.~\ref{fig:mAFSand_plot}), and can in principle be used to rule out some cases. These two observables are explicitly given as 
\begin{equation}
\sumnu = m_1 + m_2 + m_3 \quad \mbox{ and } \quad \mee = \left|U_{e1}^2 \, m_1 + U_{e2}^2 \, m_2 + U_{e3}^2 \, m_3 \right| .
\end{equation}
It is useful to plot these two quantities against each other, in both the unperturbed and perturbed case. The solid black lines in Fig.~\ref{fig:mAFSand_plot} represent the allowed ranges for normal and inverted ordering, using the best-fit values of the oscillation parameters from Table~\ref{table:oscparameters}, and varying the Majorana phases. The dotted and dashed lines include the $3\sigma$ variation in the oscillation data, for normal and inverted ordering, respectively. The scatter plots display the results of the analysis discussed above. The deviations from \ac{TBM} lead to more overlap between the normal and inverted hierarchies, with two Higgs singlets. Increasing the number of Higgs singlets effectively increases the allowed range for both $\mee$ and $\sumnu$, and one can see that the three-singlet case corresponds to the most general \ac{TBM} mass matrix [see Eq.~\eqref{eq:mtbm}]. To give one example of the consequences of Fig.~\ref{fig:mAFSand_plot}, note from the middle left panel that if $\mee$ is experimentally determined to be less than about $10^{-2}$ eV, the case with two singlets and normal mass hierarchy can be ruled out.

In general, i.e., without any model constraining the mass matrices, it is possible for $\mee$ to vanish for the normal mass hierarchy. In the case of one Higgs singlet, for example, vanishing $\mee$ means that the (1,1) entry of the mass matrix in Eq.~\eqref{eq:mnu_AF1H} (or Eq.~\eqref{eq:mnudev_AF1H} in the perturbed case) is zero, i.e., $a=2d/3$. Using the mass eigenvalues from Eq.~\eqref{eq:afdiag}, it follows that the ratio of mass-squared differences is
\begin{equation}
  r = \frac{\Delta m^2_{21}}{\Delta m^2_{31}} = \frac{2a + d}{4a} = \frac{1}{8}\ ,
\label{eq:ratio1H}
\end{equation}
which is inconsistent with the data ($r$ should be close to 1/30). Perturbing the VEV alignment [Eq.~\eqref{eq:mnudev_AF1H}] and setting the (1,1) entry of $M_\nu^{(1)'}$ to zero gives $r \simeq \frac 18 \, (1 + \epsilon_2 - 3 \epsilon_1)$, which can become sufficiently small. Indeed, in the plot $\mee$ can take values well below $10^{-2}$~eV. Similar evaluations can be made for the other cases, in this and the next section.

\section{The Altarelli-Feruglio type B seesaw model} \label{sect:AFSSmodel}

According to the classification introduced in Table~\ref{table:a4_models_assign}, type B models have lepton doublets transforming as $\ul{3}$, charged lepton singlets as $\ul{1}, \ul{1}', \ul{1}''$, and right-handed neutrinos transforming as $\ul{3}$. Neutrino mass can be generated by the type I seesaw mechanism or, when weak scalar triplets are introduced, with the type I + II seesaw mechanism. 

\subsection{The original \ac{A-F} seesaw model}

The model in Section~\ref{sect:AFmodel} can be extended by introducing right-handed neutrino fields $\nu^c$, transforming as $\ul{3}$ under $A_4$ \cite{Altarelli:2005yx}. The new Lagrangian contains all the terms in Eq.~\eqref{eq:lag_AFmodelnew}, along with the additional terms
\begin{align}
  \mathscr{L}_{\rm Y (seesaw)} &= \ y(\nu^cL)h_u +  x_A\xi(\nu^c\nu^c) + x_D(\varphi'\nu^c\nu^c)  \notag \\[2mm]
  &[+\ x_C\xi'(\nu^c\nu^c)'' + x_B\xi''(\nu^c\nu^c)'] + {\rm H.c.} + \dots \ , 
\label{eq:lag_AFss}
\end{align}
where $y$ is a coupling constant.\footnote{In Eq.~\eqref{eq:lag_AFss}
the compact notation of Eq.~\eqref{eq:lag_AFmodelnew} does not apply.}
Most details of the model, including the VEV alignment 
in Eq.~(\ref{eq:AFmodel_Higgsalign}), remain the same, with the charged lepton mass matrix given by Eq.~\eqref{eq:mlep_AF}. The Dirac mass matrix $M^D_\nu$ 
is $y v_u$ times the identity matrix, and the Majorana mass matrix is
\begin{equation}
   M_R = \begin{pmatrix} a + \frac{2d}{3} & -\frac{d}{3} & -\frac{d}{3} \\[1mm] \cdot & \frac{2d}{3} & a - \frac{d}{3} \\[1mm] \cdot & \cdot & \frac{2d}{3} \end{pmatrix}\Lambda\ ,
\end{equation}
where $a=2x_A\frac{u_a}{\Lambda}$ and $d=2x_D\frac{v'}{\Lambda}$. With the type I seesaw mechanism ($M_\nu = (M^D_\nu)^TM_R^{-1}M^D_\nu$), the light neutrino mass matrix is
\begin{equation}
  M^{(1)}_\nu = \frac{m_0}{3a(a + d)} \begin{pmatrix} 3a + d & d & d \\[1mm] \cdot & \frac{2ad+d^2}{d-a} & \frac{d^2-ad-3a^2}{d-a} \\[1mm] \cdot & \cdot & \frac{2ad+d^2}{d-a} \end{pmatrix}\ ,
\label{eq:mnu_afss}
\end{equation}
with $m_0 = y^2\frac{v_u^2}{\Lambda}$. As in Eq.~\eqref{eq:afdiag}, the matrix in Eq.~\eqref{eq:mnu_afss} is diagonalized by the \ac{TBM} matrix, with eigenvalues $m_1 = m_0/(a+d)$, $m_2 = m_0/a$ and $m_3 = m_0/(-a+d)$, leading to the sum-rule $2/m_2 + 1/m_3 = 1/m_1$. Note that the inverted hierarchy is possible in the seesaw version of this model, even with only one Higgs singlet. 

\begin{figure}[b]
  \centering
  \subfigure[Normal hierarchy]{\label{fig:ad_nh_AFSS}
  \includegraphics[width=.48\textwidth]{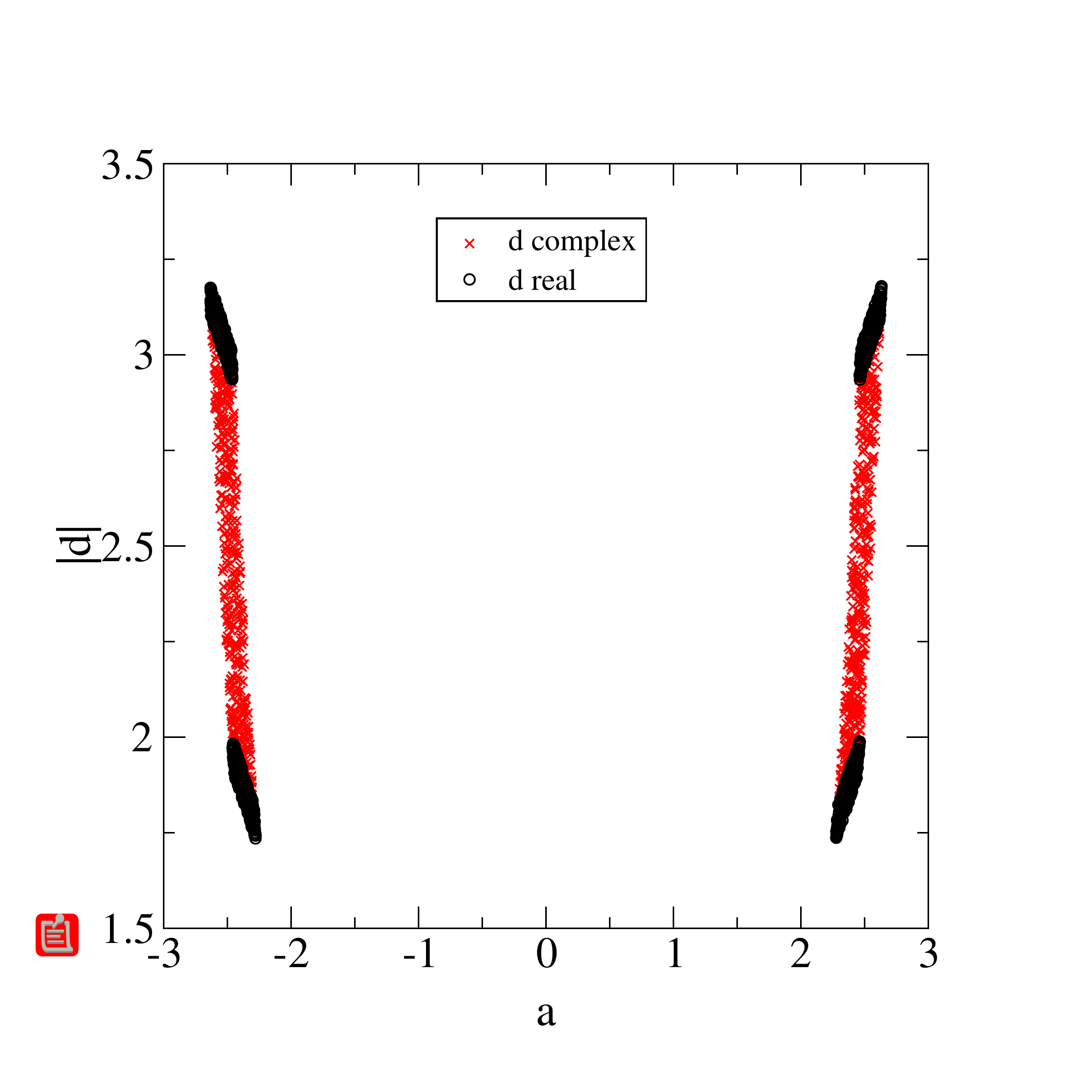}}
  \subfigure[Inverted hierarchy]{\label{fig:ad_ih_AFSS}
  \includegraphics[width=.48\textwidth]{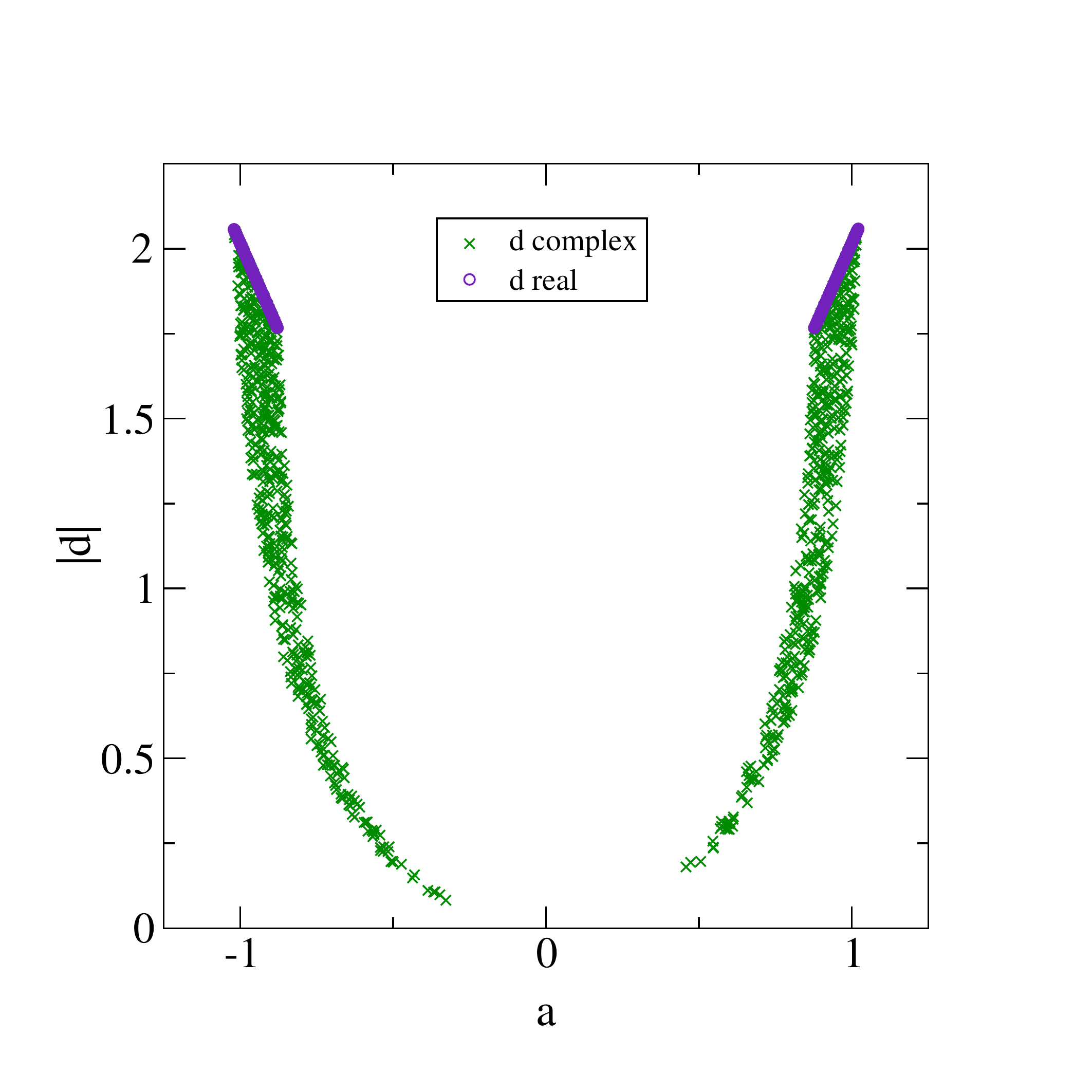}}
  \caption[Scatter plots of the $a-d$ parameter space for the A-F seesaw model, with one Higgs singlet]{Scatter plots of the $a-d$ parameter space for the A-F seesaw model, with one Higgs singlet, for normal and inverted hierarchy.}
  \label{fig:massparam_AFSS1H}
\end{figure}

The fine-tuning test again shows that in order for the correct values of the mass-squared differences to be reproduced, the parameters $a$ and $d$ must take rather specific values, 
as shown in Fig.~\ref{fig:massparam_AFSS1H}. There is a similar amount of tuning as in the A-F model without seesaw (see Fig.~\ref{fig:avb_AF1Higgs_plot}).

\subsection{Two Higgs singlets in the seesaw model}

\begin{figure}[ht]
  \centering
  \subfigure[Normal hierarchy]{\label{fig:cvd_nh_AFSS}
  \includegraphics[width=.48\textwidth]{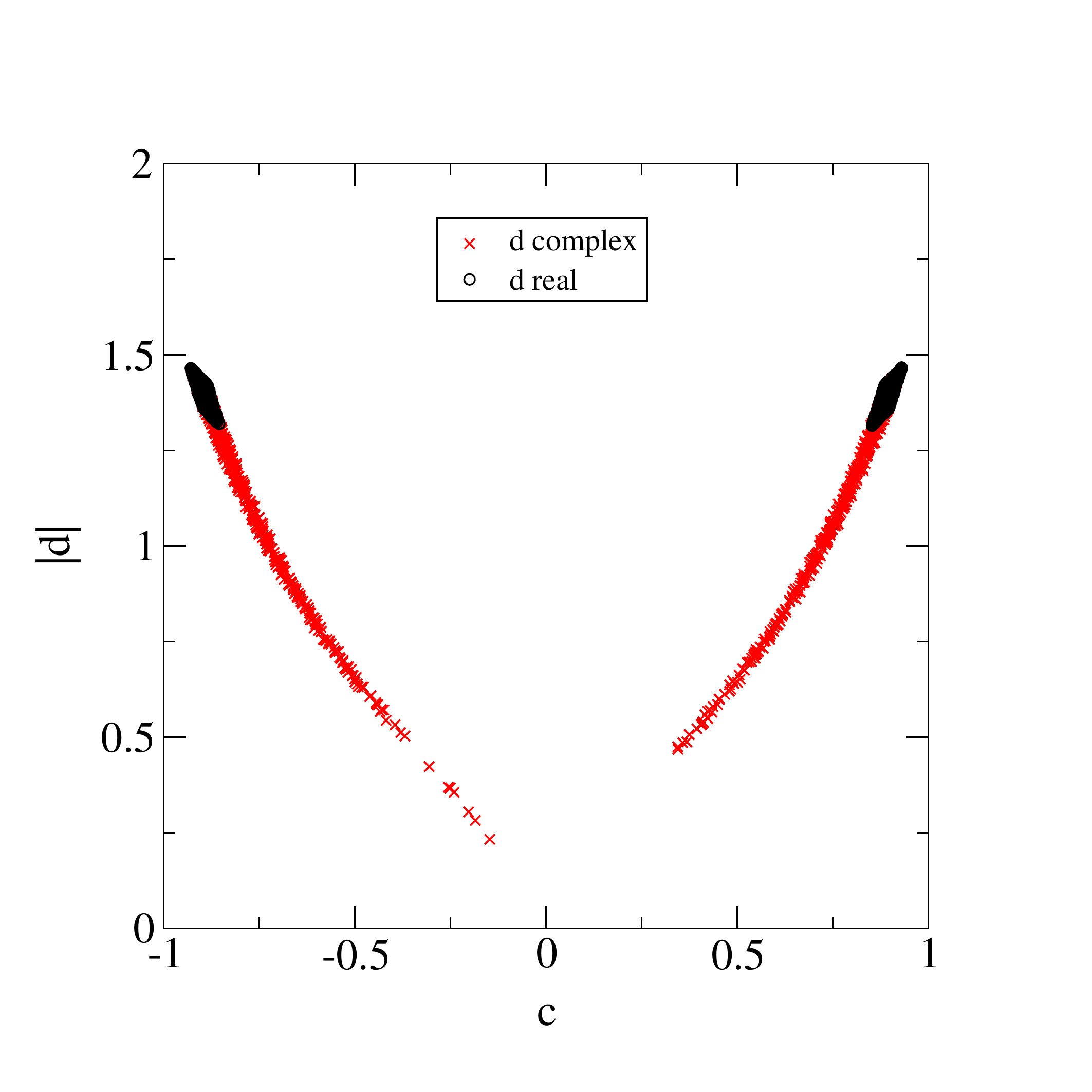}}
  \subfigure[Inverted hierarchy]{\label{fig:cvd_ih_AFSS}
  \includegraphics[width=.48\textwidth]{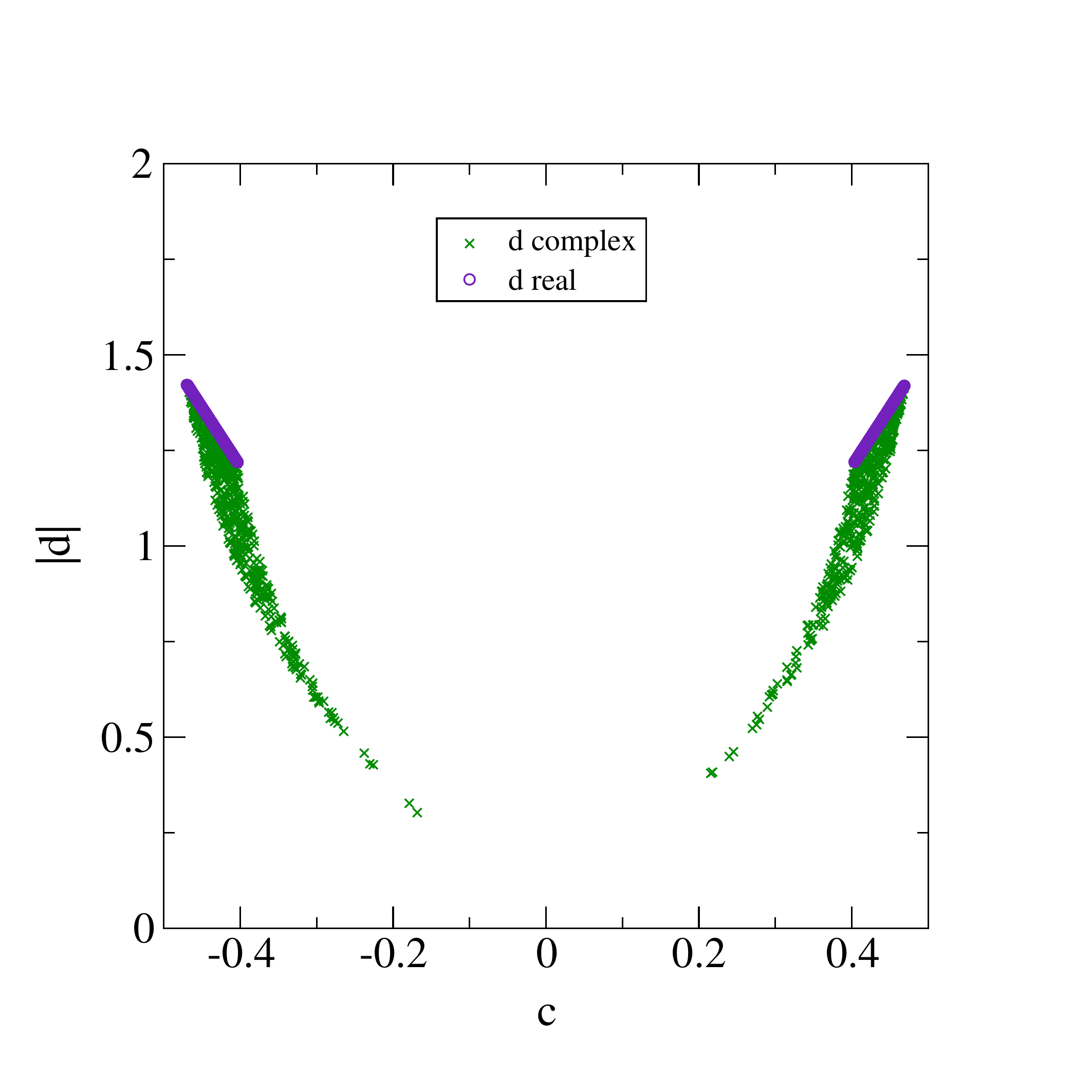}}
  \caption[Scatter plots of the $c-d$ parameter space for the A-F seesaw model, with two Higgs singlet]{Scatter plots of the $c-d$ parameter space for the A-F seesaw model, with two Higgs singlet, for normal and inverted hierarchy.}
  \label{fig:massparam_AFSS2H}
\end{figure}

In the original \ac{A-F} model, the addition of extra Higgs singlets still allows for \ac{TBM}, with certain conditions (Sections~\ref{sect:af2higgs}~and~\ref{sect:af3higgs}). This idea can also be applied to the seesaw version of the model. Again, it is possible to introduce singlets $\xi'$ and $\xi''$, transforming as $\ul{1}'$ and $\ul{1}''$, respectively. However, just like the non-seesaw case, the singlet combinations $\xi$, $\xi'$ and $\xi$, $\xi''$ cannot give rise to TBM. That is only achieved with the two singlets $\xi'$ and $\xi''$, resulting in the light neutrino mass matrix
\begin{equation}
  M^{(2)}_\nu = \frac{m_0}{m^{(2)}} \begin{pmatrix} -d^2-2 (b+c) d-3 b c & 3 b^2+d b+(c-d) d & 3 c^2+d c+(b-d) d \\[1mm] \cdot & 3 c^2-d^2-2 (b+c) d & (c-d) d+b (d-3 c) \\[1mm]  \cdot & \cdot & 3 b^2-d^2-2 (b+c) d \end{pmatrix}\ ,
\label{eq:mnu_afss2h}
\end{equation}
with $m^{(2)} = 3 \left(b^3 + c^3 - (b + c) d^2\right)$, $b=2x_B\frac{u_b}{\Lambda}$ and $c=2x_C\frac{u_c}{\Lambda}$. The condition $b=c$ is required for exact \ac{TBM}, as before, and once again the $Z_3$ symmetry means that the charged lepton sector is unaffected. The neutrino mass eigenvalues are $m_1 = m_0/(-c+d)$, $m_2 = m_0/2c$ and $m_3 = m_0/(c+d)$, and the mass sum-rule $1/m_3-1/m_1 = 1/m_2$ applies. The scatter plots in Fig.~\ref{fig:massparam_AFSS2H} show the allowed regions in $c-d$ parameter space, and exhibit a similar level of tuning as the one singlet case (Fig.~\ref{fig:massparam_AFSS1H}).

\subsection{Three Higgs singlets in the seesaw model}

\begin{figure}[ht]
  \centering
  \subfigure[Normal hierarchy]{\label{fig:abd_nh_AFSS}
  \includegraphics[width=.7\textwidth]{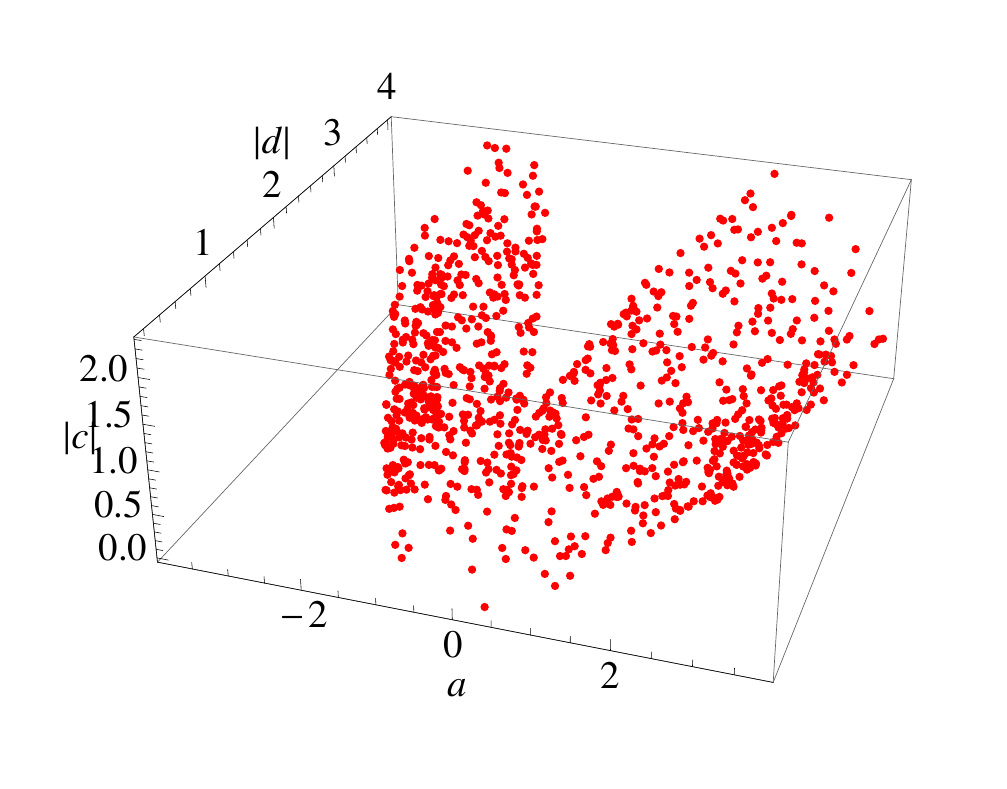}}
  \subfigure[Inverted hierarchy]{\label{fig:abd_ih_AFSS}
  \includegraphics[width=.7\textwidth]{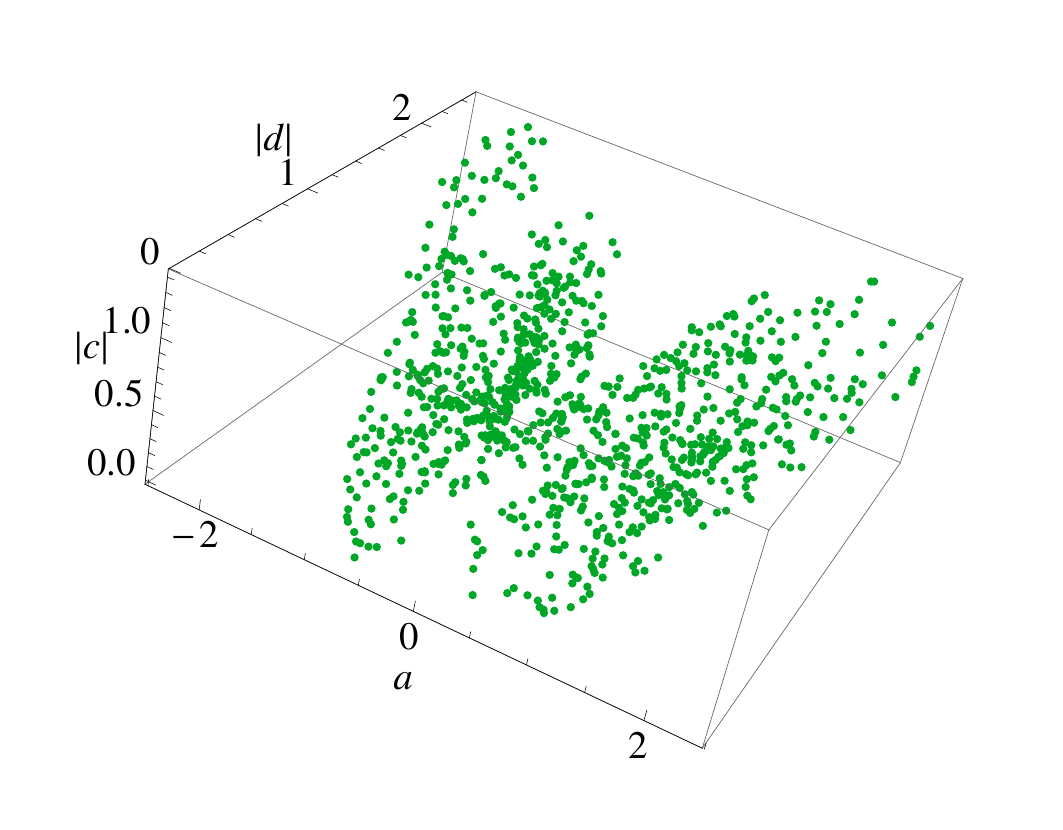}}
  \caption[Scatter plots of the $a-c-d$ parameter space for the A-F seesaw model, with three Higgs singlet]{Scatter plots of the $a-c-d$ parameter space for the A-F seesaw model, with three Higgs singlet, for normal and inverted hierarchy, with the condition $b=c$.}
  \label{fig:massparam_AFSS3H}
\end{figure}

If there are three Higgs singlets present, the light neutrino mass matrix is given by
\begin{equation}
  M^{(3)}_\nu = \frac{m_0}{m^{(3)}} M^{(3)}\ ,
\end{equation}
where the elements of the symmetric matrix $M^{(3)}$ are
\begin{align}
  M^{(3)}_{11} &= 3 a^2-d^2-3 b c-2 (a+b+c) d\ , \\
  M^{(3)}_{12} &= -d^2+(a+b+c) d+3 \left(b^2-a c\right)\ , \\
  M^{(3)}_{13} &= 3c^2+(b+c-d) d+a (d-3 b)\ , \\  
  M^{(3)}_{22} &= 3 c^2-d^2-3 a b-2 (a+b+c) d\ , \\
  M^{(3)}_{23} &= 3 a^2+d a-3 b c+(b+c-d) d\ , \\ 
  M^{(3)}_{33} &= 3 b^2-d^2-3 a c-2 (a+b+c) d\ ,
\end{align}
and $m^{(3)} = 3 \left(a^3 - 3 abc + b^3 + c^3 - (a + b + c)
d^2\right)$. Once again, the condition $a \neq b = c$ is required for
exact \ac{TBM}. In this case the neutrino mass eigenvalues become $m_1
= m_0/(a-c+d)$, $m_2 = m_0/(a+2c)$, $m_3 = m_0/(-a+c+d)$, and there is
more freedom in choosing parameters, as shown in the scatter plots of
$a-c-d$ parameter space in Fig.~\ref{fig:massparam_AFSS3H}. As in the
non-seesaw model, for three singlets hardly any tuning is
necessary. 

\subsection{Deviations from \ac{TBM} in the \ac{A-F} seesaw model}

The seesaw model can be analyzed for deviations from \ac{TBM} due to
\ac{VEV} misalignment, following the procedure outlined in
Section~\ref{sect:devaf} above, with the same limits for the
parameters. 
The VEV alignment is perturbed as in Eq.~(\ref{eq:AF_VEVdev}) and, for 
the cases of two or three singlets, as in Eq.~(\ref{eq:bc}). 
With the deviated Higgs triplet alignments of Eq.~\eqref{eq:AF_VEVdev}, the charged lepton mass matrix is again defined by Eq.~\eqref{eq:mlepdev_AF}, and the light neutrino mass matrix is
\begin{align}
    {M^{(1)}_\nu}' &= \frac{3m_0}{{m^{(1)}}'}\begin{small}\begin{pmatrix} \left((d_1-3 a)^2-4 d_2 d_3\right) & -2 d_3^2+3 a d_2-d_1 d_2 & -2 d_2^2+3 a d_3- d_1 d_3 \\
    \cdot &  \left(d_2^2-6 a d_3-4d_1 d_3\right) & 9 a^2+3 d_1 a-2 d_1^2-d_2 d_3 \\
 \cdot & \cdot & \left(d_3^2-6 a d_2-4 d_1 d_2\right) \end{pmatrix}\end{small}\ , \notag \\[2mm]
   {m^{(1)}}' &= 27 a^3-9 a \left(d_1^2+2 d_2 d_3\right)+2 \left(d_1^3-3 d_1d_2d_3+d_2^3+d_3^3\right),
\label{eq:mnu_afssdev}
\end{align}
with $d_1 = d$, $d_2 = d(1+\epsilon_1)$ and $d_3 = d(1+\epsilon_2)$. The deviated neutrino mass matrix for two Higgs singlets is
\begin{equation}
  {M^{(2)}_\nu}' = \frac{3m_0}{{m^{(2)}}'}\,{M^{(2)}}'\ ,
\label{eq:mnu_afss2hdev}
\end{equation}
where the elements of the symmetric matrix ${M^{(2)}}'$ are
\begin{align}
   {M^{(2)}_{11}}' &= d_1^2 - (3c + 2d_2)(3b + 2d_3)\ , \\
   {M^{(2)}_{12}}' &= 9b^2 + 3d_3b - 2d_3^2+ 3cd_1 - d_1d_2\ , \\
   {M^{(2)}_{13}}' &= 9c^2 + 3d_2c - 2d_2^2 + 3bd_1 - d_1d_3\ , \\ 
   {M^{(2)}_{22}}' &= (d_2 - 3c)^2 - 2d_1(3b + 2d_3)\ , \\
   {M^{(2)}_{23}}' &= -2d_1^2 - (d_2 - 3c)(d_3 - 3b)\ , \\ 
   {M^{(2)}_{33}}' &= (d_3 - 3b)^2 - 2d_1(3c + 2d_2)\ ,
\end{align}
and
\begin{align}
   {m^{(2)}}' &= -6d_1(d_2(3b + d_3) + 3cd_3) + (d_3 - 3b)^2(3b + 2d_3) \notag \\[1mm] & \quad + (d_2 - 3c)^2(3c + 2d_2) + 2d_1^3 \ ;
\end{align}
for three Higgs singlets, the deviated mass matrix is
\begin{equation}
  M'_\nu = \frac{3m_0}{{m^{(3)}}'}\,{M^{(3)}}'\ ,
\label{eq:mnu_afss3h}
\end{equation}
where the elements of ${M^{(3)}}'$ are
\begin{align}
   {M^{(3)}_{11}}' &= (d_1 - 3a)^2 - (3c + 2d_2)(3b + 2d_3)\ , \\
   {M^{(3)}_{12}}' &= 9b^2 + 3d_3b - 2d_3^2 + (d_1 - 3a) (3c - d_2)\ , \\
   {M^{(3)}_{13}}' &= 9c^2 + 3d_2c - 2d_2^2 + 3bd_1 - d_1d_3 + 3a(d_3 - 3b)\ , \\ 
   {M^{(3)}_{22}}' &= -(3a + 2d_1)(3b + 2d_3) + (d_2 - 3c)^2\ , \\
   {M^{(3)}_{23}}' &=  9a^2 + 3d_1a - 2d_1^2 + 3b(d_2 - 3c) + 3cd_3 - d_2d_3\ , \\ 
   {M^{(3)}_{33}}' &= -(3a + 2d_1)(3c + 2d_2) + (d_3 - 3b)^2 \ ,
\end{align}
and
\begin{align}
   {m^{(3)}}' &= 27 a^3-9 a \left(9 b c+d_1^2+2 d_2 d_3\right)+27 b^3-9 b \left(2d_1 d_2+d_3^2\right) \notag \\[1mm] & \quad -6 d_1 d_3 (3c+d_2)+(d_2-3 c)^2 (3 c+2 d_2)+2 d_1^3+2 d_3^3\ ,
\end{align}
with $d_1$, $d_2$ and $d_3$ as defined above. The effect of changing the relative singlet alignment is studied by setting $b = c(1+\epsilon_3)$ in both the two and three singlet cases.

\begin{figure}[ht]
  \centering
  \subfigure[Normal hierarchy]{\label{fig:sinsq1323_AFSS_n}
  \includegraphics[width=.48\textwidth]{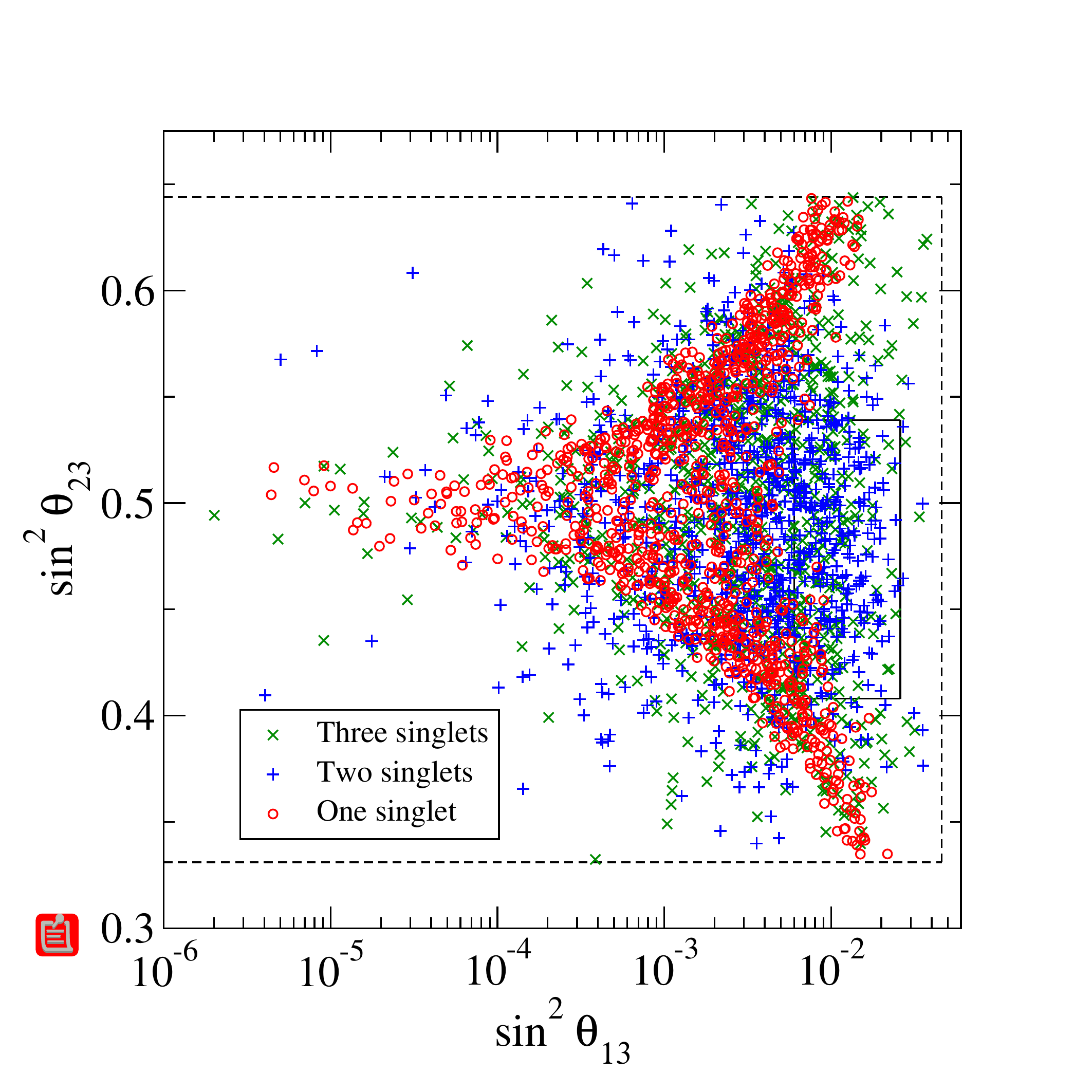}}  
  \subfigure[Inverted hierarchy]{\label{fig:sinsq1323_AFSS_i}
  \includegraphics[width=.48\textwidth]{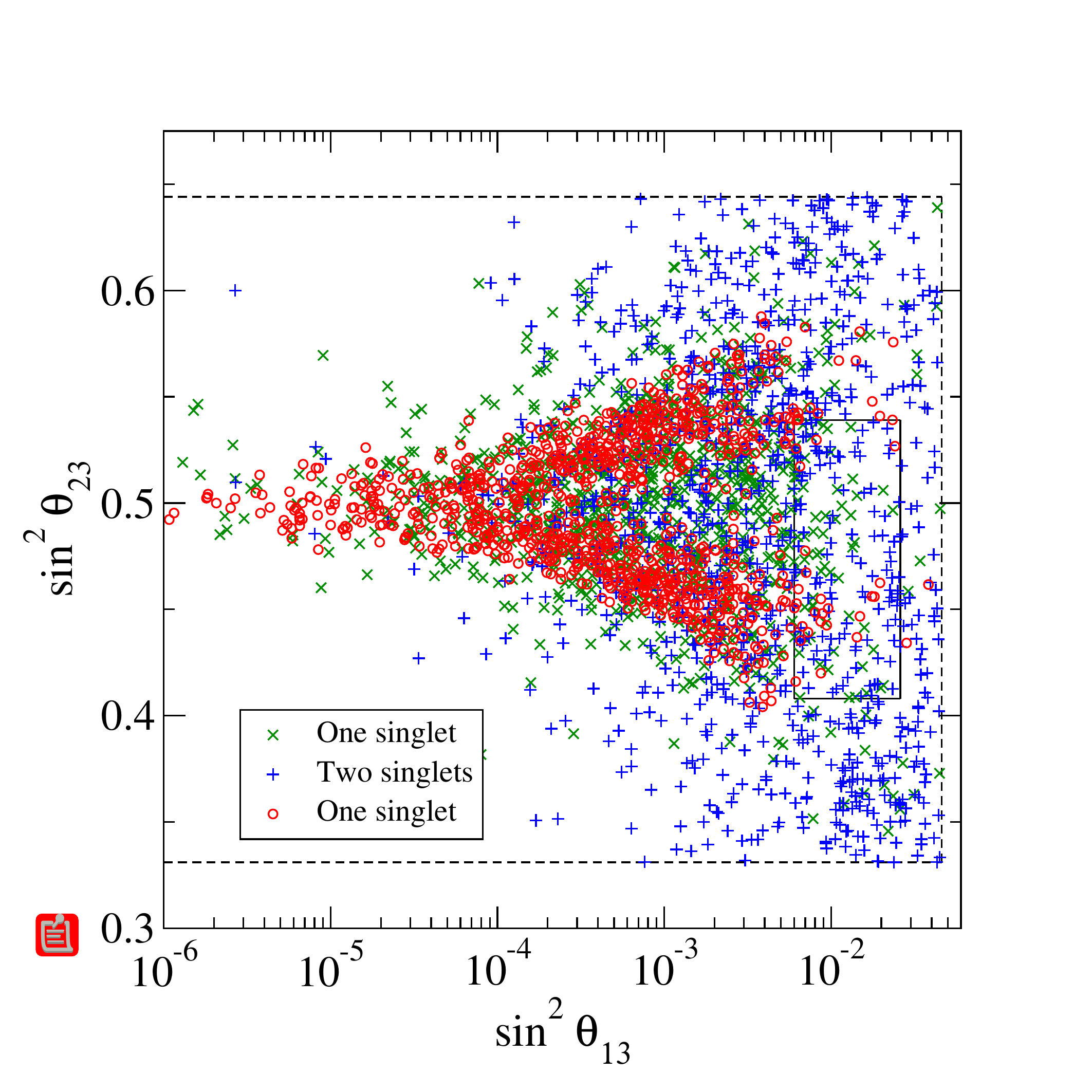}}
  \caption[Scatter plots of mixing angle observables for the A-F seesaw model, with one, two and three Higgs singlets]{Scatter plots of mixing angle observables for the A-F seesaw model \cite{Altarelli:2005yx}, with one, two and three Higgs singlets, for both the normal and inverted hierarchy.}
\label{fig:sinsq_AFSS}
\end{figure}
\begin{figure}[ht]
  \centering
  \subfigure[Normal hierarchy]{\label{fig:JCP_AFSS_n}
  \includegraphics[width=.48\textwidth]{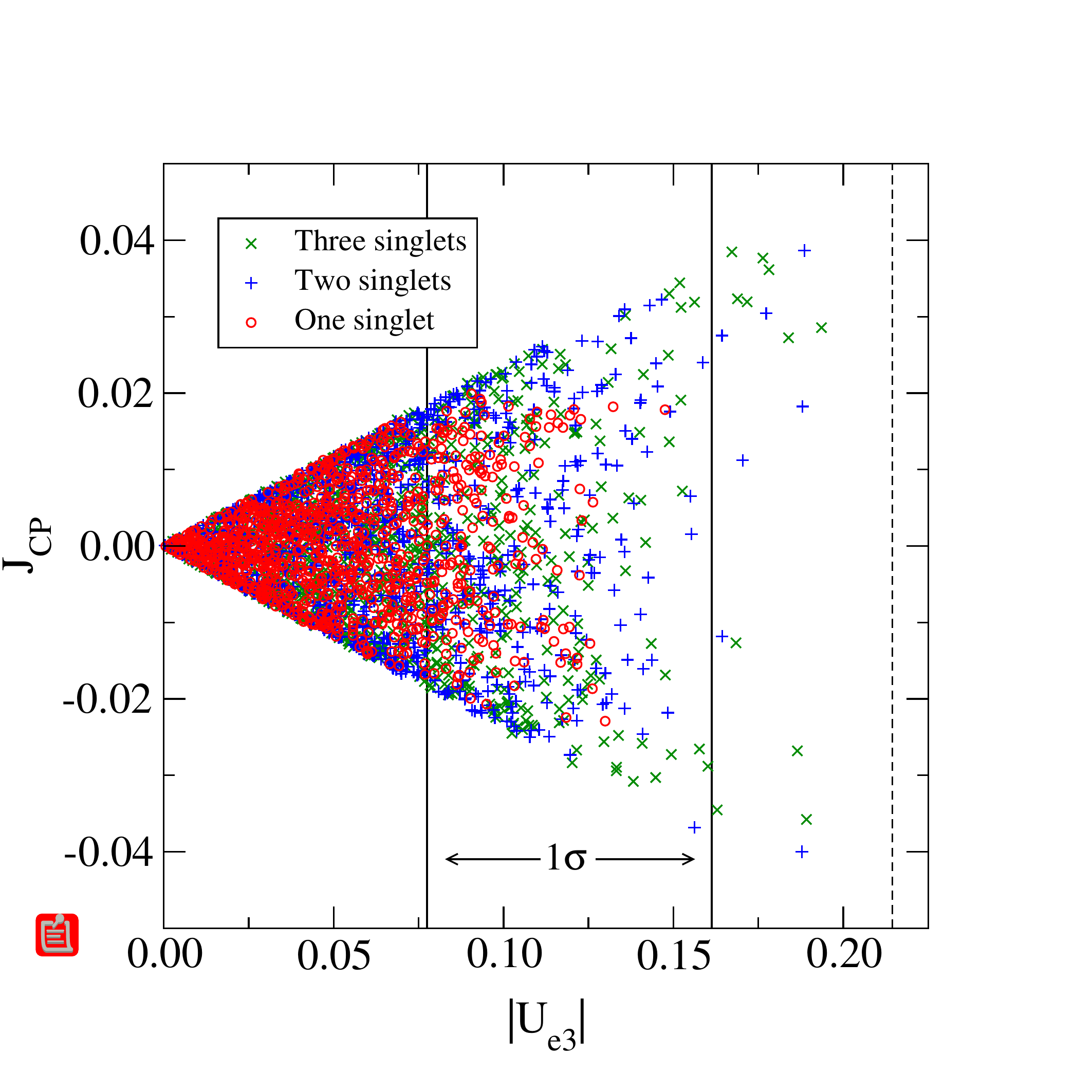}}
  \subfigure[Inverted hierarchy]{\label{fig:JCP_AFSS_i}
  \includegraphics[width=.48\textwidth]{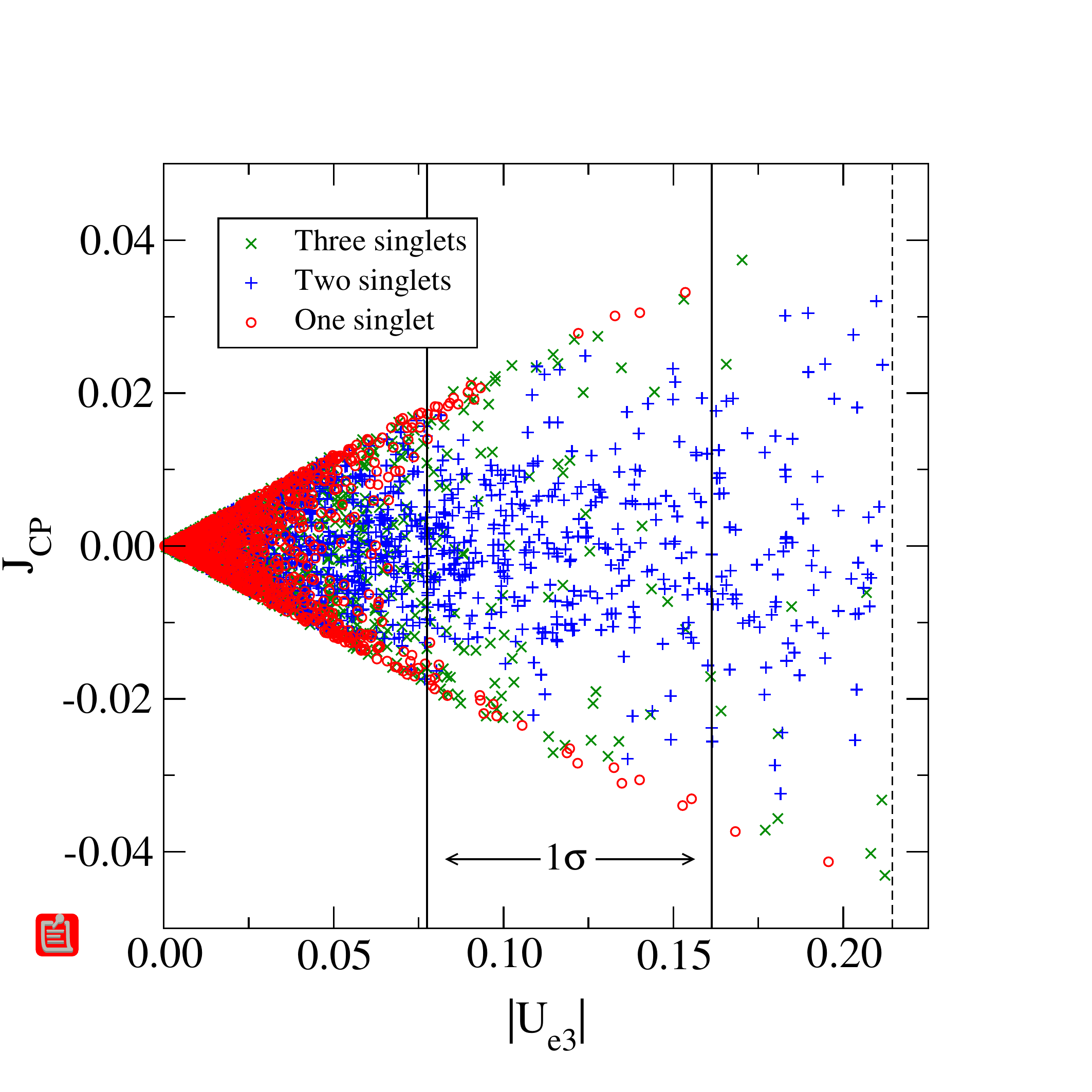}}
  \caption[Scatter plot of $J_{\rm CP}$ against $|U_{e3}|$ for the A-F seesaw model]{Scatter plot of $J_{\rm CP}$ against $|U_{e3}|$ for the A-F seesaw model, with one, two and three Higgs singlets, for both the normal and inverted hierarchy.}
  \label{fig:JCP_AFSS}
\end{figure}

Figures~\ref{fig:sinsq_AFSS} and \ref{fig:JCP_AFSS} show scatter plots 
for the mixing angles,
from diagonalization of Eqs.~\eqref{eq:mnu_afssdev},
\eqref{eq:mnu_afss2hdev} and \eqref{eq:mnu_afss3h}. 
Again, there are small differences, and in general the deviations 
from TBM can become larger with increasing number of singlets. 
However, there is little discriminative power with regards to the
number of singlets, and also with respect to the model treated in
Section \ref{sect:typeA}. 


\begin{figure}[ht]
  \centering 
  \includegraphics[width=0.8\textwidth]{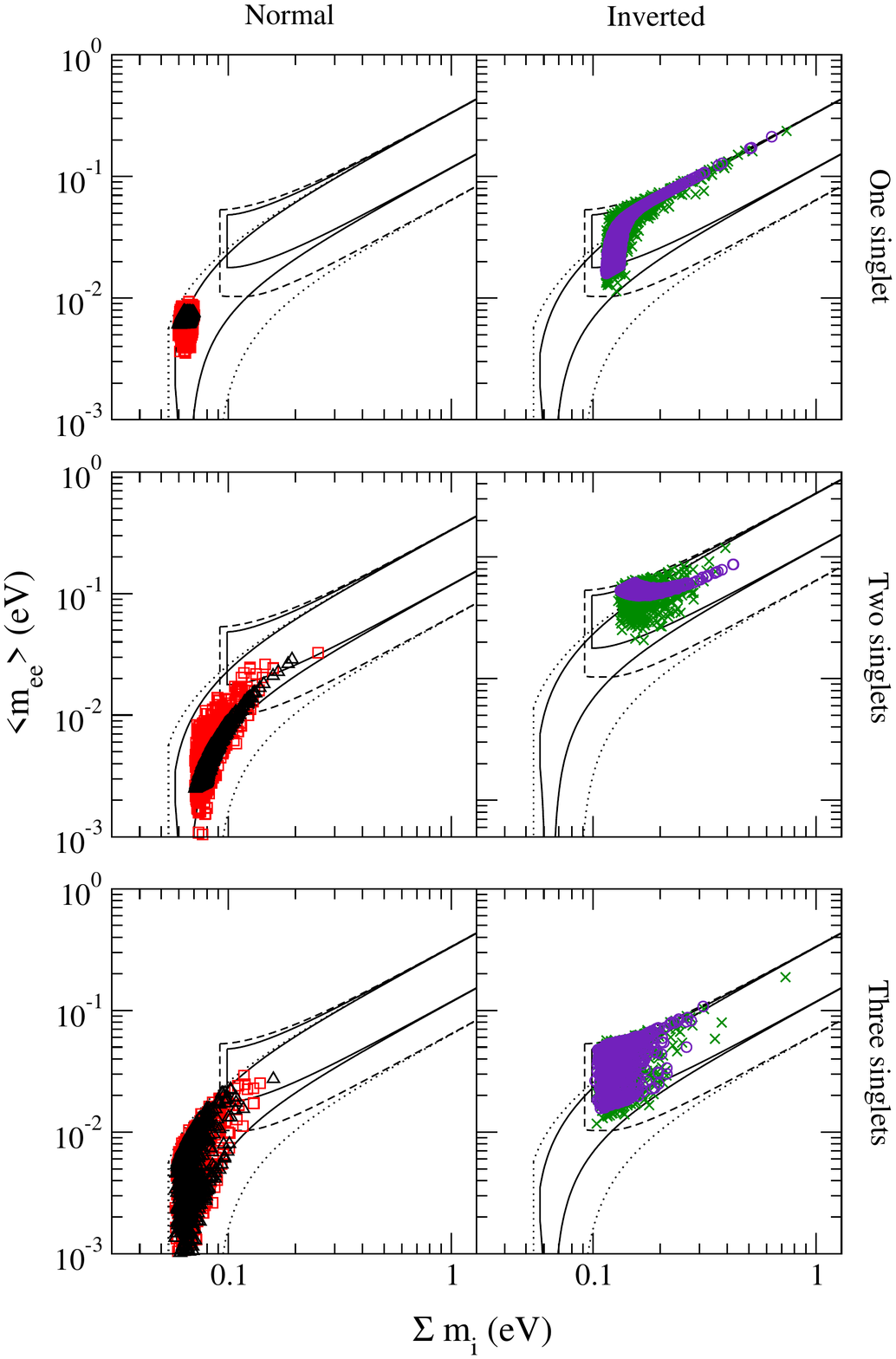}
\caption[Scatter plot of mass observables for the A-F seesaw model, with one, two and three Higgs singlets]{Scatter plot of $\mee$ against $\sumnu$ for the A-F seesaw model, with one, two and three Higgs singlets, for normal and inverted hierarchy. See the caption of Fig.~\ref{fig:mAFSand_plot} for an explanation of the symbols used.}
  \label{fig:mAFSS_plot}
\end{figure}

In spite of this, the mass dependent observables, plotted in Fig.~\ref{fig:mAFSS_plot}, allow some conclusions to be drawn. For instance, in the one singlet case there is a distinct separation of normal and inverted hierarchy, and the normal hierarchy case is very different to the non-seesaw model (Fig.~\ref{fig:mAFSand_plot}). As another example, if the normal mass hierarchy is favored by experiment and $\mee$ is measured to be 0.05 eV, the upper left panel of Fig.~\ref{fig:mAFSS_plot} shows that the seesaw model with one singlet can be ruled out. 

\section{Conclusion} \label{sect:conclusion}

The present paper is a study of deviations from TBM due to \ac{VEV}
misalignment in $A_4$ models. After an attempt to classify the vast
amount of literature according to the representations under which the
lepton doublets, lepton singlets and seesaw particles transform
under $A_4$, two particularly popular examples from classes A and B
have been focused on. The models have been checked for tuning and
then generalized, in the sense that extra singlets, transforming under
representations of $A_4$ that are not used in the original models, are
added. In general, the more singlets that are introduced, the less
tuning there is. The most general VEV misalignment is allowed for, and
the consequences for the lepton mixing observables are studied. Since
these quantities have little discriminative power, the focus is
shifted to the observables related to neutrino mass. The scatter plots
of $\mee-\sumnu$ parameter space are different in each model, and
allow one to distinguish different models, even after deviation of the
VEV alignment. This is an indication of the importance of neutrinoless
double beta decay and cosmological mass determination 
in disentangling neutrino mass models. 

\csection{Acknowledgements}
We thank the German Academic Exchange Service (DAAD) for a research grant. This work was supported by the ERC under the Starting Grant MANITOP and by the Deutsche Forschungsgemeinschaft in the Transregio 27 ``Neutrinos and beyond -- weakly interacting particles in physics, astrophysics and cosmology''. J.B.~would like to thank Steven~Karataglidis for useful discussions and advice throughout the course of this project.

\csection{Appendix}
\appendix
\renewcommand{\theequation}{A\arabic{equation}}
\setcounter{equation}{0}
\section{$A_4$ tetrahedral symmetry}

The following is an outline of the $A_4$ symmetry group \cite{Ma:2001a4,Zee:2005a4,Altarelli:2007gb}, upon which the models in this analysis are based.

\subsection{Introduction to $A_4$}

$A_4$ is the alternating group of order $4$, and is also the group of all even permutations of four objects, isomorphic to the group of rotational symmetries of the regular tetrahedron. It is a finite, non-Abelian subgroup of SO(3) \cite{Berger:2009tt} and SU(3). $A_4$ has 12 elements, which can be divided into 4 conjugacy classes with membership 1, 3, 4 and 4. The dimensionality theorem implies that there are 4 irreducible representations with dimension $d_j$ such that $\sum_jd^2_j = 12$. The only solution is $d_1 = d_2 = d_3 = 1$ and $d_4 = 3$, and the representations are labeled as $\ul{1}$, $\ul{1}'$, $\ul{1}''$ and $\ul{3}$, which means that there are three one-dimensional representations and one three-dimensional representation. The character table of $A_4$ is shown in Table~\ref{table:A4character}, with $\omega \equiv e^{i2\pi/3}$ the cube root of unity.\footnote{Note that $\omega = e^{i2\pi/3} = -1/2 + \sqrt{3}/2$ satisfies $\omega^2 = \omega^*$ and $1+\omega+\omega^2 = 0$.}
\begin{table}[ht]
\centering
\caption[Character table of $A_4$]{Character table of $A_4$, where $n$ represents the number of elements in each conjugacy class.}
\label{table:A4character}
\begin{tabular}{cccccc}
\hline \hline \T \B Class & $n$ & $\chi^{1}$ & $\chi^{1'}$ & $\chi^{1''}$ & $\chi^{3}$ \\
\hline \T $C_1$ & 1 & 1 & 1 & 1 & 3 \\
 $C_2$ & 4 & 1 & $\omega$ & $\omega^2$ & 0 \\
 $C_3$ & 4 & 1 & $\omega^2$ & $\omega$ & 0 \\
 $C_4$ & 3 & 1 & 1 & 1 & -1 \\[1.2mm]
\hline \hline
\end{tabular}
\end{table}

\subsection{Different bases for $A_4$}

There are two bases for $A_4$ commonly used in lepton family symmetry models: the \acf{M-R} basis and the \acf{A-F} basis.

\subsubsection{Ma-Rajasekaran basis} \label{subsect:MRbasis}

$A_4$ can be generated by two basic permutations $S$ and $T$, given by $S = (4321)$ and $T = (2314)$, where the generic permutation $(1,2,3,4) \rightarrow (n_1,n_2,n_3,n_4)$ is denoted by $(n_1n_2n_3n_4)$. It follows that
\begin{equation}
  S^2 = T^3 = (ST)^3 = 1\, , 
\end{equation}
which defines a ``presentation'' of the group. The one-dimensional unitary representations are generated by
\begin{eqnarray}
  \ul{1} :& S = 1 & T = 1 \ ,\notag \\
  \ul{1}' :& S = 1 & T = e^{i2\pi/3} \equiv \omega \ ,\\
  \ul{1}'' :& S = 1 & T = e^{i4\pi/3} \equiv \omega^2\ , \notag
\end{eqnarray}
and the three-dimensional unitary representation (in this basis) is built up from the generators
\begin{equation}
  S = \begin{pmatrix} 1 & 0 & 0 \\ 0 & -1 & 0 \\ 0 & 0 & -1 \end{pmatrix}\ , \quad T = \begin{pmatrix} 0 & 1 & 0 \\ 0 & 0 & 1 \\ 1 & 0 & 0 \end{pmatrix}\ .
\label{eq:af3x3gen}
\end{equation}
The $3 \times 3$ matrices of the natural three-dimensional representation $\underline{3}$ are:
\begin{align}
  C_1 &: \begin{pmatrix} 1 & 0 & 0 \\ 0 & 1 & 0 \\ 0 & 0 & 1 \end{pmatrix}, \notag \\[2mm]
  C_2 &: \begin{pmatrix} 0 & 0 & 1 \\ 1 & 0 & 0 \\ 0 & 1 & 0 \end{pmatrix}, \begin{pmatrix} 0 & 0 & 1 \\ -1 & 0 & 0 \\ 0 & -1 & 0 \end{pmatrix}, \begin{pmatrix} 0 & 0 & -1 \\
  1 & 0 & 0 \\ 0 & -1 & 0 \end{pmatrix}, \begin{pmatrix} 0 & 0 & -1 \\ -1 & 0 & 0 \\ 0 & 1 & 0 \end{pmatrix}, \notag \\[2mm]
  C_3 &: \begin{pmatrix} 0 & 1 & 0 \\ 0 & 0 & 1 \\ 1 & 0 & 0 \end{pmatrix}, \begin{pmatrix} 0 & 1 & 0 \\ 0 & 0 & -1 \\ -1 & 0 & 0 \end{pmatrix}, \begin{pmatrix} 0 & -1 & 0 \\
  0 & 0 & 1 \\ -1 & 0 & 0 \end{pmatrix}, \begin{pmatrix} 0 & -1 & 0 \\ 0 & 0 & -1 \\ 1 & 0 & 0 \end{pmatrix}, \notag \\[2mm]
  C_4 &: \begin{pmatrix} 1 & 0 & 0 \\ 0 & -1 & 0 \\ 0 & 0 & -1 \end{pmatrix}, \begin{pmatrix} -1 & 0 & 0 \\ 0 & 1 & 0 \\ 0 & 0 & -1 \end{pmatrix}, \begin{pmatrix} -1 & 0 & 0 \\
  0 & -1 & 0 \\ 0 & 0 & 1 \end{pmatrix}\ ,
\label{eq:a43x3}
\end{align}
where each matrix can be generated by $S$ and $T$ in Eq.~\eqref{eq:af3x3gen}. It is evident that the characters of the $\ul{3}$ representation (the last column of Table~\ref{table:A4character}) are simply the traces of the matrices in each class.

The multiplication rules are given by
\begin{align}
  &\ul{1} \times \ul{1} = \ul{1}\ , \label{eq:a4prod1}\\
  &\ul{1}' \times \ul{1}'' = \ul{1}\ , \\
  &\ul{1}'' \times \ul{1}' = \ul{1}\ , \\
  &\ul{1}' \times \ul{1}' = \ul{1}''\ , \\
  &\ul{1}'' \times \ul{1}'' = \ul{1}'\ , \\
  &\ul{3} \times \ul{3} = \ul{1} + \ul{1}' + {1}'' + \ul{3}_{as} + \ul{3}_s\ ,
\label{eq:a4prod6}
\end{align}
where $\ul{3}_{as}$ and $\ul{3}_s$ are ``asymmetric'' and ``symmetric'' combinations respectively. If $\ul{3}_a \sim (a_1,a_2,a_3)$ and $\ul{3}_b \sim (b_1,b_2,b_3)$ are two triplets transforming by the matrices in Eq.~\eqref{eq:a43x3}, then the three singlets and two triplets in the product in Eq.~\eqref{eq:a4prod6} are
\begin{gather}
  \ul{1} = a_1b_1 + a_2b_2 + a_3b_3\ , \label{eq:a4tripletprod1} \\
  \ul{1}' = a_1b_1 + \omega^2 a_2b_2 + \omega a_3b_3\ , \\
  \ul{1}'' = a_1b_1 + \omega a_2b_2 + \omega^2 a_3b_3\ , \\
  \ul{3}_1 \sim (a_2b_3,a_3b_1,a_1b_2)\ , \\
  \ul{3}_2 \sim (a_3b_2,a_1b_3,a_2b_1)\ .
\label{eq:a4tripletprod6}
\end{gather}

\subsubsection{Altarelli-Feruglio basis} \label{subsect:AFbasis}

In the \ac{M-R} basis, the generator $S$ in Eq.~\eqref{eq:af3x3gen} is diagonal. However, one can also represent $A_4$ in a basis where $T$ is diagonal, obtained through the unitary transformation:
\begin{align}
  T' &= V^\dagger TV = \begin{pmatrix} 1 & 0 & 0 \\ 0 & \omega & 0 \\ 0 & 0 & \omega^2 \end{pmatrix}\ , \\[1mm]
  S' &= V^\dagger SV = \frac{1}{3} \begin{pmatrix} -1 & 2 & 2 \\ 2 & -1 & 2 \\ 2 & 2 & -1 \end{pmatrix}\ ,
\end{align}
where
\begin{equation}
  V = \frac{1}{\sqrt{3}} \begin{pmatrix} 1 & 1 & 1 \\ 1 & \omega^2 & \omega \\ 1 & \omega & \omega^2 \end{pmatrix}\ .
\label{eq:magicm}
\end{equation}
It is known that the most general mass matrix leading to TBM, 
\begin{equation}
m_\nu^{\rm TBM} = 
\begin{pmatrix}
A & B & B \\
\cdot & \frac12 (A + B + D) & \frac12 (A + B - D) \\
\cdot & \cdot & \frac12 (A + B + D)
\end{pmatrix}
\label{eq:mtbm}
\end{equation}
is invariant with respect to $S'$: $(S')^T 
\, m_\nu^{\rm TBM} S' = m_\nu^{\rm TBM}$. 
Note that the matrix $V$ is the so-called ``magic matrix'', which
appears in some $A_4$ models as the unitary matrix that diagonalizes the charged lepton mass matrix. In the $S'$, $T'$ basis, the multiplication rules are identical to those in Eqs.~\eqref{eq:a4prod1} -- \eqref{eq:a4prod6}, but the product of two triplets gives the composition of the following irreducible representations:
\begin{gather}
  \ul{1} = a_1b_1 + a_2b_3 + a_3b_2\ , \\
  \ul{1}' = a_3b_3 + a_1b_2 + a_2b_1\ , \\
  \ul{1}'' = a_2b_2 + a_1b_3 + a_3b_1\ , \\
  \ul{3}_s \sim \frac{1}{3}\left(2a_1b_1 - a_2b_3 - a_3b_2\ , 2a_3b_3 - a_1b_2 - a_2b_1\ , 2a_2b_2 - a_1b_3 - a_3b_1\right)\ , \\
  \ul{3}_{as} \sim \frac{1}{3}\left(a_2b_3 - a_3b_2, a_1b_2 - a_2b_1, a_1b_3 - a_3b_1\right)\ .
\label{eq:a4tripletprod2}
\end{gather}

\subsection{Equivalence of the two bases}

The model presented in Section~\ref{sect:AFmodel} can be formulated in the \ac{M-R} basis, using the same particle assignments and the Lagrangian in Eq.~\eqref{eq:lag_AFmodelnew}. With the product decomposition rules in Eqs.~\eqref{eq:a4tripletprod1} -- \eqref{eq:a4tripletprod6}, and the triplet \ac{VEV} alignment
\begin{equation}
  \langle\varphi\rangle = (v,v,v) \quad {\rm and} \quad \langle\varphi'\rangle = (v',0,0)\ ,
\label{eq:vevs_alt}
\end{equation}
the charged lepton and neutrino mass matrices are 
\begin{equation}
  M_\ell = v_d\frac{v}{\Lambda} \begin{pmatrix} y_e & y_e & y_e \\ y_\mu & y_\mu\omega^2 & y_\mu\omega \\ y_\tau & y_\tau\omega & y_\tau\omega^2 \end{pmatrix}\ , \quad M_\nu = \frac{v_u}{\Lambda} \begin{pmatrix} a & 0 & 0 \\ \cdot & a & d \\ \cdot & \cdot & a \end{pmatrix}\ .
\label{eq:massm_alt}
\end{equation}
In this case, $M_\ell$ is diagonalized by the magic matrix
[Eq.~\eqref{eq:magicm}], and $M_\nu$ is diagonalized by
\begin{equation}
  V_\nu = \begin{pmatrix} 0 & 1 & 0 \\ \frac{1}{\sqrt{2}} & 0 & -\frac{1}{\sqrt{2}} \\ \frac{1}{\sqrt{2}} & 0 & \frac{1}{\sqrt{2}} \end{pmatrix}\ , 
\end{equation}
which combines with $V$ in Eq.~\eqref{eq:magicm} to give $U_{\rm TBM}$. The neutrino mass matrix in Eq.~\eqref{eq:massm_alt} is equivalent to that in Eq.~\eqref{eq:mnu_AF1H}, with the change of basis induced by $V$. Thus the two bases lead to equivalent models, with the triplet \ac{VEV} alignments in the charged lepton and neutrino sectors effectively swapped [compare Eqs.~\eqref{eq:AFmodel_Higgsalign} and \eqref{eq:vevs_alt}]. Note that the change of basis will change the relative phases of the eigenvalues of $M_\nu$.

\bibliographystyle{h-physrev}
\bibliography{Masters}

\begin{thebibliography}{10}

\bibitem{Fogli:2005cq}
G.~L. Fogli, E.~Lisi, A.~Marrone, and A.~Palazzo,
\newblock Prog. Part. Nucl. Phys. {\bf 57}, 742 (2006), hep-ph/0506083.

\bibitem{Fogli:2006yq}
G.~L. Fogli {\em et~al.},
\newblock Phys. Rev. D {\bf 75}, 053001 (2007), hep-ph/0608060.

\bibitem{Fogli:2008ig}
G.~L. Fogli {\em et~al.},
\newblock Phys. Rev. D {\bf 78}, 033010 (2008), 0805.2517.

\bibitem{Schwetz:2008osc}
T.~Schwetz, M.~Tortola, and J.~W.~F. Valle,
\newblock New J. Phys. {\bf 10}, 113011 (2008), 0808.2016.

\bibitem{Harrison:2002tbm}
P.~F. Harrison, D.~H. Perkins, and W.~G. Scott,
\newblock Phys. Lett. {\bf B530}, 167 (2002), hep-ph/0202074.

\bibitem{Pakvasa:2007zj}
S.~Pakvasa, W.~Rodejohann, and T.~J. Weiler,
\newblock Phys. Rev. Lett. {\bf 100}, 111801 (2008), 0711.0052.

\bibitem{Altarelli:2010gt}
G.~Altarelli and F.~Feruglio,
\newblock (2010), 1002.0211.

\bibitem{Altarelli:2005a4}
G.~Altarelli and F.~Feruglio,
\newblock Nucl. Phys. {\bf B720}, 64 (2005), hep-ph/0504165.

\bibitem{Zee:2005a4}
A.~Zee,
\newblock Phys. Lett. {\bf B630}, 58 (2005), hep-ph/0508278.

\bibitem{Adhikary:2006wi}
B.~Adhikary, B.~Brahmachari, A.~Ghosal, E.~Ma, and M.~K. Parida,
\newblock Phys. Lett. {\bf B638}, 345 (2006), hep-ph/0603059.

\bibitem{Altarelli:2005yx}
G.~Altarelli and F.~Feruglio,
\newblock Nucl. Phys. {\bf B741}, 215 (2006), hep-ph/0512103.

\bibitem{Honda:2008rs}
M.~Honda and M.~Tanimoto,
\newblock Prog. Theor. Phys. {\bf 119}, 583 (2008), 0801.0181.

\bibitem{Brahmachari:2008a4}
B.~Brahmachari, S.~Choubey, and M.~Mitra,
\newblock Phys. Rev. D {\bf 77}, 073008 (2008), 0801.3554.

\bibitem{Feruglio:2008ht}
F.~Feruglio, C.~Hagedorn, Y.~Lin, and L.~Merlo,
\newblock Nucl. Phys. {\bf B809}, 218 (2009), 0807.3160.

\bibitem{Morisi:2009qa}
S.~Morisi,
\newblock Phys. Rev. D {\bf 79}, 033008 (2009), 0901.1080.

\bibitem{Morisi:2009sz}
S.~Morisi,
\newblock J. Phys. Conf. Ser. {\bf 203}, 012060 (2010), 0910.2542.

\bibitem{Altarelli:2006kg}
G.~Altarelli, F.~Feruglio, and Y.~Lin,
\newblock Nucl. Phys. {\bf B775}, 31 (2007), hep-ph/0610165.

\bibitem{Bazzocchi:2007na}
F.~Bazzocchi, S.~Kaneko, and S.~Morisi,
\newblock JHEP {\bf 03}, 063 (2008), 0707.3032.

\bibitem{Ma:2004zv}
E.~Ma,
\newblock Phys. Rev. D {\bf 70}, 031901 (2004), hep-ph/0404199.

\bibitem{Ma:2005a42}
E.~Ma,
\newblock Phys. Rev. D {\bf 72}, 037301 (2005), hep-ph/0505209.

\bibitem{Ma:2002yp}
E.~Ma,
\newblock Mod. Phys. Lett. {\bf A17}, 627 (2002), hep-ph/0203238.

\bibitem{Babu:2002dz}
K.~S. Babu, E.~Ma, and J.~W.~F. Valle,
\newblock Phys. Lett. {\bf B552}, 207 (2003), hep-ph/0206292.

\bibitem{Hirsch:2003dr}
M.~Hirsch, J.~C. Romao, S.~Skadhauge, J.~W.~F. Valle, and A.~Villanova~del
  Moral,
\newblock Phys. Rev. D {\bf 69}, 093006 (2004), hep-ph/0312265.

\bibitem{He:2006dk}
X.-G. He, Y.-Y. Keum, and R.~R. Volkas,
\newblock JHEP {\bf 04}, 039 (2006), hep-ph/0601001.

\bibitem{Altarelli:2008bg}
G.~Altarelli, F.~Feruglio, and C.~Hagedorn,
\newblock JHEP {\bf 03}, 052 (2008), 0802.0090.

\bibitem{Burrows:2009pi}
T.~J. Burrows and S.~F. King,
\newblock Nucl. Phys. {\bf B835}, 174 (2010), 0909.1433.

\bibitem{Ma:2001a4}
E.~Ma and G.~Rajasekaran,
\newblock Phys. Rev. D {\bf 64}, 113012 (2001), hep-ph/0106291.

\bibitem{Babu:2005se}
K.~S. Babu and X.-G. He,
\newblock (2005), hep-ph/0507217.

\bibitem{Ma:2005qf}
E.~Ma,
\newblock Phys. Rev. D {\bf 73}, 057304 (2006), hep-ph/0511133.

\bibitem{Yin:2007rv}
F.~Yin,
\newblock Phys. Rev. D {\bf 75}, 073010 (2007), 0704.3827.

\bibitem{Adhikary:2008au}
B.~Adhikary and A.~Ghosal,
\newblock Phys. Rev. D {\bf 78}, 073007 (2008), 0803.3582.

\bibitem{Csaki:2008qq}
C.~Csaki, C.~Delaunay, C.~Grojean, and Y.~Grossman,
\newblock JHEP {\bf 10}, 055 (2008), 0806.0356.

\bibitem{Chen:2009um}
M.-C. Chen and S.~F. King,
\newblock JHEP {\bf 06}, 072 (2009), 0903.0125.

\bibitem{Hayakawa:2009va}
A.~Hayakawa, H.~Ishimori, Y.~Shimizu, and M.~Tanimoto,
\newblock Phys. Lett. {\bf B680}, 334 (2009), 0904.3820.

\bibitem{Berger:2009tt}
J.~Berger and Y.~Grossman,
\newblock JHEP {\bf 02}, 071 (2010), 0910.4392.

\bibitem{Ding:2009gh}
G.-J. Ding and J.-F. Liu,
\newblock JHEP {\bf 05}, 029 (2010), 0911.4799.

\bibitem{Mitra:2009jj}
M.~Mitra,
\newblock (2009), 0912.5291.

\bibitem{delAguila:2010vg}
F.~del Aguila, A.~Carmona, and J.~Santiago,
\newblock (2010), 1001.5151.

\bibitem{Dong:2010gk}
P.~V. Dong, L.~T. Hue, H.~N. Long, and D.~V. Soa,
\newblock Phys. Rev. {\bf D81}, 053004 (2010), 1001.4625.

\bibitem{Ma:2006vq}
E.~Ma,
\newblock Mod. Phys. Lett. {\bf A22}, 101 (2007), hep-ph/0610342.

\bibitem{Ma:2006wm}
E.~Ma,
\newblock Mod. Phys. Lett. {\bf A21}, 2931 (2006), hep-ph/0607190.

\bibitem{Bazzocchi:2007au}
F.~Bazzocchi, S.~Morisi, and M.~Picariello,
\newblock Phys. Lett. {\bf B659}, 628 (2008), 0710.2928.

\bibitem{Ma:2009wi}
E.~Ma,
\newblock (2009), 0908.3165.

\bibitem{Hirsch:2005mc}
M.~Hirsch, A.~Villanova~del Moral, J.~W.~F. Valle, and E.~Ma,
\newblock Phys. Rev. D {\bf 72}, 091301 (2005), hep-ph/0507148.

\bibitem{Morisi:2007ft}
S.~Morisi, M.~Picariello, and E.~Torrente-Lujan,
\newblock Phys. Rev. D {\bf 75}, 075015 (2007), hep-ph/0702034.

\bibitem{Ciafaloni:2009ub}
P.~Ciafaloni, M.~Picariello, E.~Torrente-Lujan, and A.~Urbano,
\newblock Phys. Rev. D {\bf 79}, 116010 (2009), 0901.2236.

\bibitem{Hirsch:2008rp}
M.~Hirsch, S.~Morisi, and J.~W.~F. Valle,
\newblock Phys. Rev. D {\bf 78}, 093007 (2008), 0804.1521.

\bibitem{Morisi:2009sc}
S.~Morisi and E.~Peinado,
\newblock Phys. Rev. {\bf D80}, 113011 (2009), 0910.4389.

\bibitem{Chen:2005jm}
S.-L. Chen, M.~Frigerio, and E.~Ma,
\newblock Nucl. Phys. {\bf B724}, 423 (2005), hep-ph/0504181.

\bibitem{Bazzocchi:2008sp}
F.~Bazzocchi, M.~Frigerio, and S.~Morisi,
\newblock Phys. Rev. D {\bf 78}, 116018 (2008), 0809.3573.

\bibitem{Bazzocchi:2008rz}
F.~Bazzocchi, S.~Morisi, M.~Picariello, and E.~Torrente-Lujan,
\newblock J. Phys. G {\bf 36}, 015002 (2009), 0802.1693.

\bibitem{Ciafaloni:2009qs}
P.~Ciafaloni, M.~Picariello, A.~Urbano, and E.~Torrente-Lujan,
\newblock Phys. Rev. {\bf D81}, 016004 (2010), 0909.2553.

\bibitem{Ma:2005a4}
E.~Ma,
\newblock Mod. Phys. Lett. {\bf A20}, 2601 (2005), hep-ph/0508099.

\bibitem{Lavoura:2006hb}
L.~Lavoura and H.~Kuhbock,
\newblock Mod. Phys. Lett. {\bf A22}, 181 (2007), hep-ph/0610050.

\bibitem{Hirsch:2007kh}
M.~Hirsch, A.~S. Joshipura, S.~Kaneko, and J.~W.~F. Valle,
\newblock Phys. Rev. Lett. {\bf 99}, 151802 (2007), hep-ph/0703046.

\bibitem{Frampton:2008ci}
P.~H. Frampton and S.~Matsuzaki,
\newblock (2008), 0806.4592.

\bibitem{Lin:2008aj}
Y.~Lin,
\newblock Nucl. Phys. {\bf B813}, 91 (2009), 0804.2867.

\bibitem{King:2006np}
S.~F. King and M.~Malinsky,
\newblock Phys. Lett. {\bf B645}, 351 (2007), hep-ph/0610250.

\bibitem{Altarelli:2009kr}
G.~Altarelli and D.~Meloni,
\newblock J. Phys. G {\bf 36}, 085005 (2009), 0905.0620.

\bibitem{Lin:2009bw}
Y.~Lin,
\newblock Nucl. Phys. {\bf B824}, 95 (2010), 0905.3534.

\bibitem{Morisi:2008nk}
S.~Morisi,
\newblock Nuovo Cim. {\bf 123B}, 886 (2008), 0807.4013.

\bibitem{Hirsch:2009mx}
M.~Hirsch, S.~Morisi, and J.~W.~F. Valle,
\newblock Phys. Lett. {\bf B679}, 454 (2009), 0905.3056.

\bibitem{Ibanez:2009du}
D.~Ibanez, S.~Morisi, and J.~W.~F. Valle,
\newblock Phys. Rev. {\bf D80}, 053015 (2009), 0907.3109.

\bibitem{Ding:2009iy}
G.-J. Ding,
\newblock Nucl. Phys. {\bf B827}, 82 (2010), 0909.2210.

\bibitem{Bazzocchi:2009pv}
F.~Bazzocchi, L.~Merlo, and S.~Morisi,
\newblock Nucl. Phys. {\bf B816}, 204 (2009), 0901.2086.

\bibitem{Plentinger:2005kx}
F.~Plentinger and W.~Rodejohann,
\newblock Phys. Lett. {\bf B625}, 264 (2005), hep-ph/0507143.

\bibitem{Dighe:2006sr}
A.~Dighe, S.~Goswami, and W.~Rodejohann,
\newblock Phys. Rev. D {\bf 75}, 073023 (2007), hep-ph/0612328.

\bibitem{Hirsch:2006je}
M.~Hirsch, E.~Ma, J.~C. Romao, J.~W.~F. Valle, and A.~Villanova~del Moral,
\newblock Phys. Rev. {\bf D75}, 053006 (2007), hep-ph/0606082.

\bibitem{Hochmuth:2007wq}
K.~A. Hochmuth, S.~T. Petcov, and W.~Rodejohann,
\newblock Phys. Lett. {\bf B654}, 177 (2007), 0706.2975.

\bibitem{Albright:2008tbm}
C.~H. Albright and W.~Rodejohann,
\newblock Phys. Lett. {\bf B665}, 378 (2008), 0804.4581.

\bibitem{Goswami:2009yy}
S.~Goswami, S.~T. Petcov, S.~Ray, and W.~Rodejohann,
\newblock Phys. Rev. D {\bf 80}, 053013 (2009), 0907.2869.

\bibitem{Li:2009kx}
Y.~F. Li and Q.~Y. Liu,
\newblock Mod. Phys. Lett. {\bf A25}, 63 (2010), 0911.2670.

\bibitem{Ge:2010js}
S.-F. Ge, H.-J. He, and F.-R. Yin,
\newblock JCAP {\bf 1005}, 017 (2010), 1001.0940.

\bibitem{Altarelli:2007gb}
G.~Altarelli,
\newblock (2007), 0711.0161.

\end{thebibliography}

\end{document}